\documentclass[letterpaper,twocolumn,10pt]{article}
\usepackage{usenix-2020-09}

\usepackage{color}
\usepackage{amsmath}
\definecolor{revisionpurple}{rgb}{0.45,0.15,0.65}

\usepackage{enumitem}
\usepackage{booktabs}
\usepackage{multirow}
\usepackage{array}
\usepackage{colortbl}
\usepackage{graphicx}
\usepackage{svg}
\usepackage{xcolor}
\usepackage{tikz}
\usetikzlibrary{positioning,fit,backgrounds,calc,trees,arrows.meta}
\usepackage{rotating}
\usepackage{forest}
\usepackage[capitalize]{cleveref}
\usepackage{tabularx}
\usepackage[most]{tcolorbox}
\usetikzlibrary{positioning,arrows.meta}
\definecolor{cVec}{HTML}{2C5F8A}     
\definecolor{cVecBg}{HTML}{E3ECF4}   
\definecolor{cRisk}{HTML}{9C3B2E}    
\definecolor{cRiskBg}{HTML}{F4E4E1}  
\definecolor{cCore}{HTML}{D9E6D4}    
\definecolor{cNeutral}{HTML}{4A4A4A} 
\definecolor{cEdge}{HTML}{8A9BA8}    
\definecolor{cAgent}{HTML}{EAEFF3}   
\definecolor{cAttack}{HTML}{6B5A8E}  
\definecolor{cAttackBg}{HTML}{EEEAF4}
\definecolor{cDefense}{HTML}{2F7D68} 
\definecolor{cDefenseBg}{HTML}{E3F0EB}
\definecolor{covHi}{HTML}{8CB87E}    
\definecolor{covMid}{HTML}{EFB66F} 
\definecolor{covLo}{HTML}{D89B92}    
\tikzset{
  fnode/.style={draw=cNeutral, line width=0.4pt, rounded corners=2pt},
  mascomponent/.style={fnode, fill=cAgent, align=center, font=\scriptsize,
                       minimum height=0.62cm, inner sep=3pt},
  masexternal/.style={mascomponent, text width=1.42cm, minimum height=1.18cm},
  masworldpart/.style={fnode, draw=cNeutral, fill=white, align=left,
                       font=\tiny, text width=1.95cm, minimum height=0.72cm,
                       inner sep=2.5pt},
  masagent/.style={fnode, draw=cVec, fill=white, align=center,
                   font=\tiny, text width=1.46cm, minimum height=1.04cm,
                   inner sep=2pt},
  masadmission/.style={mascomponent, draw=cVec, fill=cCore,
                       text width=1.35cm, minimum height=0.78cm},
  masgateway/.style={mascomponent, draw=cRisk, fill=cRiskBg,
                     font=\tiny, text width=2.1cm, minimum height=0.82cm},
  mascommit/.style={mascomponent, draw=cVec, fill=cCore,
                   text width=1.58cm, minimum height=0.9cm},
  masworldbox/.style={fnode, draw=cNeutral, fill=white, inner sep=5pt},
  massociety/.style={fnode, draw=cVec, fill=white, inner sep=5pt},
  masprincipal/.style={mascomponent, draw=cVec, fill=cCore, text width=1.28cm},
  maspeer/.style={mascomponent, draw=cVec, fill=white, text width=1.12cm,
                  minimum height=0.46cm},
  masstate/.style={mascomponent, text width=2.25cm, minimum height=0.72cm},
  mascapability/.style={mascomponent, text width=1.55cm, minimum height=0.7cm},
  masoutcome/.style={mascomponent, text width=1.52cm, minimum height=1.02cm},
  masplane/.style={fnode, draw=cNeutral, fill=white, align=center,
                   font=\tiny, text width=5.35cm, minimum height=0.46cm},
  masboundary/.style={fnode, draw=cNeutral, fill=cVecBg!24, inner sep=7pt},
  masiface/.style={fnode, draw=cVec, fill=cVecBg, text=cVec, align=center,
                   font=\tiny, text width=1.3cm, minimum height=0.43cm,
                   inner sep=2pt},
  masflow/.style={->, draw=cEdge, line width=0.72pt},
  masstateflow/.style={<->, draw=cEdge, line width=0.65pt},
  masauthority/.style={->, draw=cRisk, dashed, line width=0.75pt},
  mastelemetry/.style={<->, draw=cEdge, densely dotted, line width=0.55pt},
  masinterfacelead/.style={draw=cVec, line width=0.35pt},
  masattack/.style={fnode, draw=cAttack, fill=cAttackBg, text=cAttack,
                    align=center, font=\tiny\bfseries, inner sep=2.5pt},
  masattackpath/.style={->, draw=cAttack, densely dashed, line width=0.95pt},
  masrisk/.style={fnode, draw=cRisk, fill=cRiskBg, text=cRisk, align=center,
                  font=\tiny\bfseries, inner sep=2.2pt},
  masrisklead/.style={draw=cRisk, line width=0.45pt},
  masdefense/.style={fnode, draw=cDefense, double=cDefenseBg,
                     double distance=0.7pt, fill=cDefenseBg, text=cDefense,
                     align=center, font=\tiny\bfseries, inner sep=2.5pt},
  masdefenselead/.style={->, draw=cDefense, line width=0.65pt},
}

\newtcolorbox{benchmarksummary}{
  width=\dimexpr\linewidth-3pt\relax,
  before={\par\noindent\hspace{3pt}},
  after={\par},
  colback=black!7,
  colframe=black,
  boxrule=0.5pt,
  arc=2mm,
  left=5pt,
  right=5pt,
  top=4pt,
  bottom=4pt,
  before skip=7pt,
  after skip=7pt,
  fontupper=\small
}

\newcommand{\paragraphtitle}[1]{\paragraph{#1}}

\title{\Large\bf SoK: When Safe Agents Fail Together:\\
The Security of Multi Agent LLM Systems}

\author{Rui Yang¹, Junjie Xu², Zhengyu Liu¹, Neil Fendley¹, Yang Hong¹, Ziyang Li¹, Yinzhi Cao¹\\
¹ Johns Hopkins University
² Nanyang Technological University\\
ryang54@jh.edu, andrexu736269@gmail.com, zliu192@jhu.edu,\\
neil.fendley@jhuapl.edu, yhong51@jh.edu, ziyang@cs.jhu.edu,
yinzhi.cao@jhu.edu}
\begin{document}

\maketitle

\begin{abstract}
Safe agents can fail together. Multi-agent LLM systems (MAS) move information, state, decisions, and authority across principal boundaries, creating failures that local checks may miss. 
Without an execution-level view, a multi-agent setting can easily be mistaken for evidence of a genuinely multi-agent security effect.
We thus systematize MAS security through an execution-centered analysis of 197 works, covering six interaction interfaces, four adversary positions, seven system-level risks, and eight recurring attack paths. We introduce an A-I-R framework that organizes attacks by adversary position, interaction interface, and resulting system-level risk, unifying otherwise fragmented attack mechanisms across MAS. We organize defenses through a five-part contract covering path target, observation, intervention, trust boundary, and recovery, and identify path closure and recovery as key challenges. 
{{We audit 44 evaluation and benchmark works and identify open challenges in isolating interaction effects, designing comparable and diagnostic metrics, supporting reuse across MAS designs, and evaluating open-system operation.}}
Together, these findings motivate an interaction-aware view of MAS security: trace attacks end to end, test whether defenses close those paths, and evaluate system-level effects with appropriate counterfactuals.
\end{abstract}

\section{Introduction}

Agents that are safe when used individually can fail when used together in a MAS.
Benign prompt fragments can become harmful when combined~\cite{arif2026conjunctive}; truthful reports can steer a group toward a false belief~\cite{hu2026lyingtruths}; and attacker-controlled content can pass through honest specialists to a privileged tool~\cite{jha2026controlflowhijacking}. These failures arise through interactions among the separately addressable participants of the MAS, which we call \emph{principals}. Messages propagate influence between principals, a shared state preserves propagated influence, aggregation combines local outputs from different principals, and delegation transfers authority across boundaries~\cite{lee2026promptinfection,liu2026topologymemory,liu2025manipulatecollective,jha2026controlflowhijacking}. We therefore study the end-to-end execution {of MAS}, where interactions may leave a failure largely unchanged, amplify it, create a failure through composition, or require defining a security property that exists only when principals interact.


\begin{figure*}[!t]
    \centering
    \includegraphics[width=0.95\linewidth]{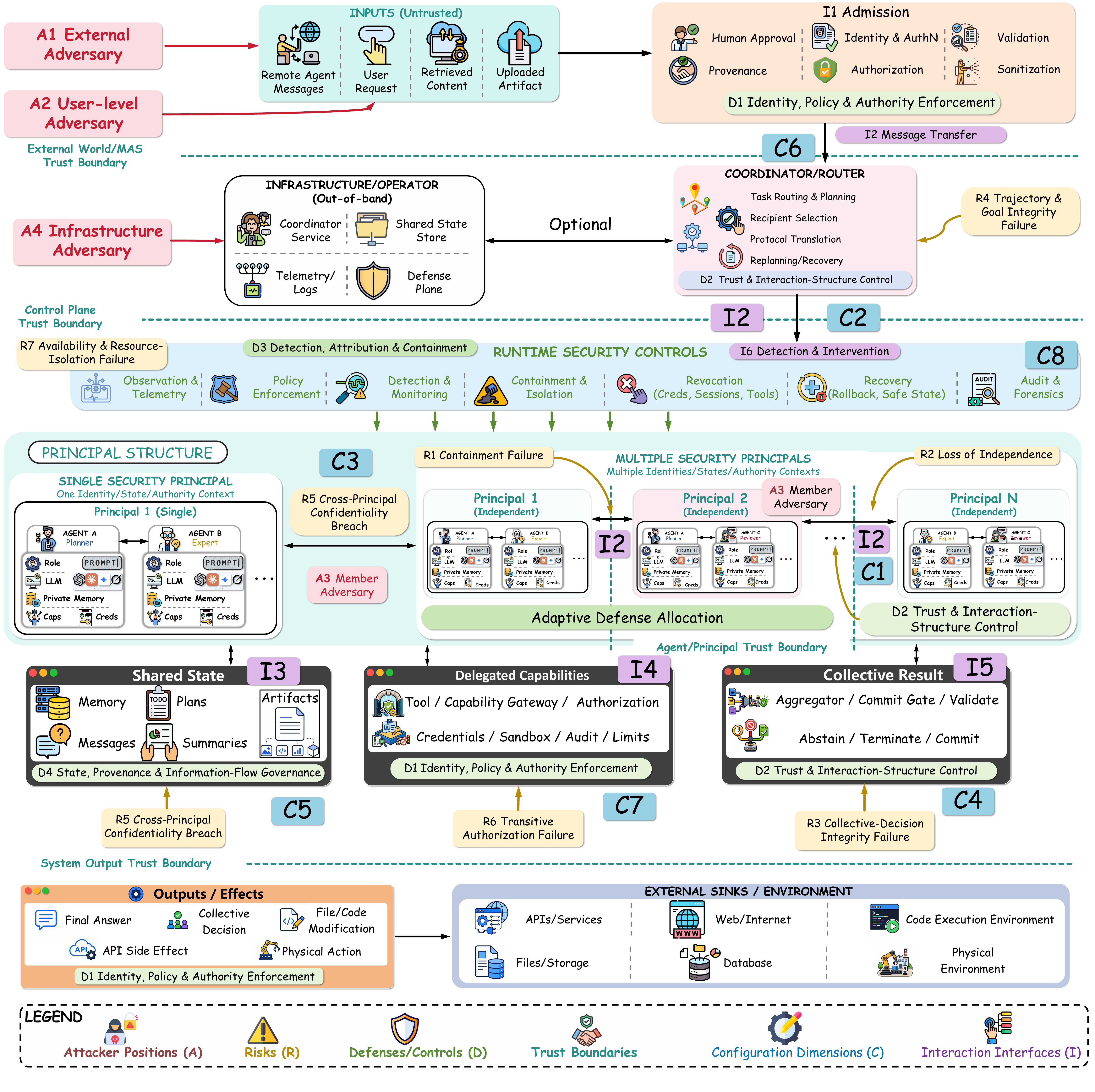}
    \caption{
    Unified execution view of MAS security.
    C1--C8 condition the execution.
    A1--A4 identify where adversarial influence begins, I1--I6 identify the
    transitions it crosses, and R1--R7 identify the resulting system-level failure.
    D1--D4 denote defensive control primitives applied along the execution.
    }
    \label{fig:overview}
\end{figure*}

Existing surveys cover many important threats, approaching from  {different perspectives}. Agent-security surveys organize risks around models, prompts, memory, tools, workflows, interfaces, and environments~\cite{deng2025agentsunderthreat,he2025emergedsecurity,yu2025trustworthyagents,kim2026agenticai}, while MAS-focused surveys emphasize communication, trust, topology, governance, and collective behavior~\cite{hammond2025multiagentrisks,schroederdewitt2025openchallenges,ko2026sevenchallenges,raza2026trism}. What remains unclear is how an attack moves through a multi-agent system: where it enters, which interaction {interfaces} it uses, and what system-level security failure it finally causes. Without this execution view, simply observing a failure in a multi-agent setting can be mistaken for evidence that the failure depends on interaction. This distinction is increasingly important as the field grows, because closely related studies still use different threat models, units of analysis, metrics, and evidence standards. A common way to describe and evaluate these security effects is therefore needed.

We address this gap by treating the complete multi-agent execution as the unit
of analysis. Our core representation traces adversarial influence from its
starting position (A), through one or more interaction interfaces
(I), to a system-level security failure (R), yielding the
\textbf{$A\rightarrow I\rightarrow R$} framework. Eight configuration
dimensions (C1--C8) describe the conditions under which such an
execution occurs; they condition a path. Across the
corpus, we group recurring $A\rightarrow I\rightarrow R$ executions into eight
non-exclusive attack paths (P1--P8). These paths are synthesis
shorthands for recurring execution patterns, not an additional taxonomy of
security outcomes. The same execution view is then used to analyze where
defenses can intervene and how security claims should be evaluated.
\Cref{fig:overview} summarizes this organization. Our contributions are therefore fourfold:


\begin{itemize}
    \item
    We define multi-agent security around end-to-end execution, providing a common basis for deciding when interaction matters to a security claim.
    This execution-centered view provides a common model for connecting system configuration, interaction paths, security effects, defenses, and evaluation.

    \item
    We introduce the $A\rightarrow I\rightarrow R$ framework and synthesize seven system-level risk families and eight recurring attack paths. This structure follows adversarial influence from entry to consequence while showing how multiple risks and paths can compose within one execution.

    \item
    We organize defenses with a five-part contract
    and map where each control primitive can interrupt an attack path. This exposes gaps in path closure, observer adequacy, intervention timing, mediation, and recovery.

    \item
    {{We audit 44 evaluation and benchmark works and identify open problems in interaction-specific comparisons, designing comparable and diagnostic metrics, benchmark portability, and open-system evaluation.}}
\end{itemize}
\section{Overview}\label{sec:overview}

This section defines the scope of the SoK, positions our analysis relative to
prior surveys, and summarizes the evidence base used for the synthesis.

\subsection{Scope}
\label{sec:scope}

\noindent\textbf{System boundary.}
We define an MAS by separately addressable principals: at least two participants
whose outputs can affect one another through messages, shared state, collective
decisions, or delegated actions
~\cite{yang2026amongus,huang2025faultyagents}. We call each such participant a
\emph{principal}. Principals may share the same LLM backend or use multiple
models internally, so the system boundary is defined by interaction rather than
model count~\cite{yang2026amongus,kim2026agenticai}. Debate systems,
orchestrated teams, swarms, hierarchies, and A2A networks are therefore in scope
when their participants act as separate principals
~\cite{gu2024agentsmith,
li2026a2asecbench,motwani2024secretcollusion}. Routing, ensembling, repeated
sampling, and planner--critic--verifier pipelines are out of scope when all
steps remain within one principal~\cite{kim2026agenticai}.

\noindent\textbf{Security boundary.}
For the synthesis, we treat a result as evidence of a MAS-specific security
effect when interaction changes a security-relevant outcome or when the
protected property itself depends on interaction. Prompt injection, for example,
may remain local, propagate across principals, persist in shared state, influence
a collective decision, or gain authority through delegation
~\cite{lee2026promptinfection,liu2026topologymemory,
liu2025manipulatecollective,jha2026controlflowhijacking}. Authentication and
authorization are likewise interaction-level concerns when trust or authority
crosses principal boundaries
~\cite{shi2025trustauthorization,xu2026trustparadox,
jha2026controlflowhijacking}. Changes in accuracy, agreement, or behavior,
including behavioral drift, count as security failures only when they violate a
stated protected property or boundary. Results that reproduce an agent
vulnerability in an MAS without a material interaction effect may be
retained as comparative context, but they do not by themselves establish a
MAS-specific causal claim.

Interaction can play four roles in a security result. An
\underline{\emph{inherited} effect} remains largely unchanged when the relevant
relation is removed; we use this as a comparison label, not evidence of
interaction dependence. An \underline{\emph{interaction-amplified} effect}
remains possible, but interaction changes likelihood, reach, persistence,
privilege, or impact
~\cite{hagag2026architecturesecurity,yu2025netsafe,
liu2026topologymemory,tian2023evilgeniuses}. A \underline{\emph{composition-induced} effect} appears
only when principals or steps are combined
~\cite{arif2026conjunctive,s1ragfailures2026,arxiv_2509_14284}. A
\underline{\emph{structurally multi-agent} effect} concerns a property defined
by relations among principals, such as collusion, coalition formation, or quorum
manipulation
~\cite{motwani2024secretcollusion,mathew2025hidden,yang2026amongus}.
For inherited, amplified, and composition-induced effects, relation-removal
counterfactuals can distinguish whether the effect persists, changes in
magnitude, or disappears when the protected property remains meaningful.
Structurally multi-agent effects require comparisons that preserve the relation.
These labels describe how interaction contributes to a security effect and what
evidence is needed to support that claim.

\begin{table}[t]
\centering
\scriptsize

\newcommand{\colWork}{1.6cm}
\newcommand{\colInteraction}{1.4cm}
\newcommand{\colTrace}{1.25cm}
\newcommand{\colThreat}{1.35cm}
\newcommand{\colDefense}{1.2cm}
\newcommand{\colEvidence}{1.45cm}

\newcommand{\covYes}{%
  \tikz[baseline=-0.55ex]{
    \fill[covHi] (0,0) circle (0.78ex);
    \draw[black!25,line width=.22pt] (0,0) circle (0.78ex);
  }%
}

\newcommand{\covHalf}{%
  \tikz[baseline=-0.55ex]{
    \fill[covMid] (0,0) circle (0.78ex);
    \begin{scope}
      \clip (0,-0.8ex) rectangle (0.8ex,0.8ex);
      \fill[white] (0,0) circle (0.78ex);
    \end{scope}
    \draw[black!30,line width=.22pt] (0,0) circle (0.78ex);
  }%
}

\newcommand{\covNo}{%
  \tikz[baseline=-0.55ex]{
    \fill[white] (0,0) circle (0.78ex);
    \draw[black!35,line width=.25pt] (0,0) circle (0.78ex);
  }%
}

\newcolumntype{L}[1]{>{\raggedright\arraybackslash}p{#1}}
\newcolumntype{C}[1]{>{\centering\arraybackslash}p{#1}}

\setlength{\tabcolsep}{0.1pt}
\renewcommand{\arraystretch}{1.15}

\begin{tabular}{@{}
  L{\colWork}
  C{\colInteraction}
  C{\colTrace}
  C{\colThreat}
  C{\colDefense}
  C{\colEvidence}
@{}}

\toprule

\textbf{Work}
&
\shortstack{\textbf{Interaction}\\\textbf{counterfactual}}
&
\shortstack{\textbf{End-to-end}\\\textbf{execution}\\\textbf{trace}}
&
\shortstack{\textbf{Threat--path--}\\\textbf{protected}\\\textbf{effect}}
&
\shortstack{\textbf{Defense}\\\textbf{assumptions}}
&
\shortstack{\textbf{Outcome-stage}\\\textbf{evidence}}
\\

\midrule

Hammond et al.~\cite{hammond2025multiagentrisks}
& \covHalf & \covNo & \covHalf & \covNo & \covNo
\\

\rowcolor{black!2.5}
Schroeder de Witt et al.~\cite{schroederdewitt2025openchallenges}
& \covHalf & \covHalf & \covHalf & \covHalf & \covNo
\\

Sun et al.~\cite{sun2026unique}
& \covHalf & \covHalf & \covHalf & \covHalf & \covHalf
\\

\rowcolor{black!2.5}
Ko et al.~\cite{ko2026sevenchallenges}
& \covHalf & \covHalf & \covHalf & \covHalf & \covHalf
\\

Raza et al.~\cite{raza2026trism}
& \covNo & \covNo & \covHalf & \covHalf & \covNo
\\

\rowcolor{black!2.5}
Rafe et al.~\cite{rafe2026orchestration}
& \covNo & \covNo & \covHalf & \covHalf & \covNo
\\

Vangalapat et al.~\cite{vangalapat2026trustworthy}
& \covNo & \covHalf & \covHalf & \covHalf & \covNo
\\

\rowcolor{black!2.5}
Kale et al.~\cite{kale2026privacyrag}
& \covNo & \covNo & \covHalf & \covHalf & \covNo
\\

Tanveer~\cite{tanveer2026orchestration}
& \covNo & \covHalf & \covHalf & \covHalf & \covNo
\\

\rowcolor{black!2.5}
Kong et al.~\cite{kong2025agentcomm}
& \covHalf & \covHalf & \covHalf & \covHalf & \covNo
\\

Reid et al.~\cite{reid2025governedrisk}
& \covHalf & \covHalf & \covHalf & \covNo & \covHalf
\\

\rowcolor{black!2.5}
Shi et al.~\cite{shi2025trustauthorization}
& \covNo & \covHalf & \covHalf & \covHalf & \covNo
\\

Nguyen et al.~\cite{nguyen2026securityconsiderations}
& \covNo & \covNo & \covHalf & \covNo & \covNo
\\

\midrule

\rowcolor{cCore!65}
\textbf{Ours}
& \covYes & \covYes & \covYes & \covYes & \covYes
\\

\bottomrule
\end{tabular}

\vspace{1pt}

{\scriptsize
\covYes\ Systematic
\hspace{4em}
\covHalf\ Partial
\hspace{4em}
\covNo\ Not systematically treated
}

\normalsize
\caption{Coverage of prior multi-agent security syntheses across our analytical dimensions, using the coding criteria stated in \cref{sub:related}.}
\label{tab:related}
\end{table}

\subsection{Related Work}
\label{sub:related}

Prior surveys, reviews, and synthesis-style studies cover complementary parts of
agent and multi-agent security. Broad agent-security surveys organize threats
and defenses around models, prompts, memory, tools, workflows, interfaces, and
environments
~\cite{deng2025agentsunderthreat,he2025emergedsecurity,
kim2026agenticai,yu2025trustworthyagents}.
MAS-focused syntheses study communication, trust, topology, coordination,
authorization, governance, and system-level risks
~\cite{hammond2025multiagentrisks,schroederdewitt2025openchallenges,
sun2026unique,ko2026sevenchallenges,kong2025agentcomm,
shi2025trustauthorization,raza2026trism}.
More focused or adjacent studies examine orchestration and verification,
trustworthy coordination, privacy-preserving retrieval, governed risk analysis,
and the coverage of existing security frameworks
~\cite{rafe2026orchestration,vangalapat2026trustworthy,
kale2026privacyrag,tanveer2026orchestration,
reid2025governedrisk,nguyen2026securityconsiderations}.

These works have different goals and scopes, and we do not treat the comparison
as a ranking of their quality. Our distinction is the unit of analysis. Rather
than organizing security primarily by component, attack name, design pattern, or
governance concern, we organize it around cross-principal execution. The same
execution model connects where adversarial influence enters, which interaction
interfaces it crosses, which protected system-level effect it reaches, where a
defense can intervene and under what assumptions, and what evidence supports the
claim. \Cref{tab:related} compares prior syntheses along these analytical
dimensions. We code a dimension as \emph{systematic} only when it is used as an
explicit organizing or evaluation axis, as \emph{partial} when it is
substantively discussed or illustrated but not applied systematically, and as
\emph{not systematically treated} otherwise. These markers describe analytical
coverage, not the quality of the cited work.

\noindent\textbf{Literature Selection.}
We assembled the corpus through searches across major security, AI, and NLP
venues and scholarly indexes, supplemented by backward and forward citation
chasing. After deduplication and review against \cref{sec:scope}, the final
corpus contains 197 works.
We divide the corpus into two evidence sets with different roles in the
synthesis. Set~1 contains 115 works that were peer reviewed or had received at
least ten citations by the frozen review cutoff. We use these works to
characterize the current MAS-security landscape and to support the main
synthesis in the paper. Set~2 contains 82 additional in-scope works that point
to emerging mechanisms, system settings, and security questions that may shape
the next stage of the field. We use these works more selectively, primarily to
highlight extensions of established attack paths, defenses, and evaluation
settings.
The literature corpus and associated artifacts will be released publicly with the final version of the paper.

\section{System Model and Interaction Surface}
\label{sec:setup}\label{sub:systemmodel}\label{def:mas-bg}

Building on the scope above, we separate the conditions around an execution
from the transitions within it. Configuration dimensions (C1--C8)
describe how the system is arranged and therefore condition the reach,
persistence, and impact of an attack. Interaction interfaces
(I1--I6) identify the security-relevant transitions through which
information, state, decisions, authority, or security controls cross boundaries.
The interfaces later form the middle of the $A\rightarrow I\rightarrow R$
execution path; the configuration dimensions provide context for interpreting
that path.

\subsection{Configuration Dimensions}
\label{sec:dimensions}

The eight configuration dimensions describe the conditions that shape an MAS
execution. The same attack may therefore have different reach, persistence, or
impact across configurations, and several dimensions may matter at once.

\noindent\textbf{C1: Communication topology.}
Communication topology describes which principals can communicate directly and
which paths connect them. MAS studies commonly consider complete, chain or line,
ring or circle, tree, and star topologies, as well as other system-specific
communication graphs
~\cite{hagag2026architecturesecurity,yu2025netsafe,wu2026cia}. Topology changes
path length, connectivity, and centrality, shaping how far adversarial influence
can travel and which principals become important intermediaries.

\noindent\textbf{C2: Protocol semantics.}
Protocol semantics describe how principals exchange information and what
security context is preserved across each exchange. MAS communication ranges
from free-form natural-language messages to messages with fixed roles or formats
and agent-to-agent protocols
~\cite{li2026a2asecbench,he2025communicationattacks,
jha2026controlflowhijacking}. Security depends on whether identity, origin and
provenance, task context, and authorization information remain available across
principal boundaries.
End-to-end enforcement may require the security context needed by downstream checks to remain available across the protocol chain.
Each exchange may be valid on its own, yet the full execution can still become
unsafe if translation, summarization, state updates, or delegation lose the
origin, purpose, provenance, or authority needed by a later decision
~\cite{jha2026controlflowhijacking,s1cascadinginstruction2026}.

\noindent\textbf{C3: Principal composition.}
Principal composition describes which principals participate, what roles they
play, and how independent they are. Systems may use homogeneous or heterogeneous
models and generalist or specialist roles. Shared models, prompts, or evidence
sources can create correlated behavior even when several principals participate
~\cite{hagag2026architecturesecurity,mathew2025hidden}.

\noindent\textbf{C4: Coordination mechanism.}
Coordination mechanisms describe how principals combine their contributions into
a system-level decision or action. Common mechanisms include debate, voting,
judge or aggregator models, consensus, and orchestrator-mediated coordination
~\cite{liu2025manipulatecollective,motwani2024secretcollusion,
hu2026lyingtruths}. Their security depends on how malicious or correlated
contributions affect the collective outcome.

\noindent\textbf{C5: State architecture.}
State architecture describes where state is stored, who can access it, and how
long it persists. Distinctions include principal-local versus shared state, ephemeral
versus persistent state, and whether state is copied or inherited across
principals
~\cite{liu2026topologymemory,agentworm2026,childinherits2026}. These choices
determine which principals can read or modify state and whether derived state can
later be traced, revoked, or repaired.

\noindent\textbf{C6: Membership and trust.}
Membership and trust describe which principals may participate and which
relationships the system treats as trusted. Membership may be closed or open,
static or dynamic, with persistent or ephemeral identities
~\cite{li2026a2asecbench,childinherits2026,supp_dynatrust_2026}. Dynamic systems
must update admission, trust, and revocation as principals join, leave, or change
roles. Membership is part of the system's security state. In open or
self-spawning systems, the population can change during execution, so the
fraction of malicious principals is not necessarily fixed. Robustness claims
based on that fraction therefore depend on how admission, identity, spawning,
and revocation define who counts as a participant
~\cite{childinherits2026,supp_dynatrust_2026}.

\noindent\textbf{C7: Authority placement.}
Authority placement describes which principals hold or can delegate privileged
capabilities. Relevant authority includes tool permissions, credentials,
resource budgets, delegation rights, and routing or control-plane privileges
~\cite{jha2026controlflowhijacking,xu2026trustparadox,
s1cascadinginstruction2026}. Security impact depends not only on which
principals are reachable, but also on what authority is reachable through them.

\noindent\textbf{C8: Oversight architecture.}
Oversight architecture describes who can observe an execution and how much of it
they can see. Monitoring may be local, distributed across principals, or
centralized with graph or trace visibility. Broader visibility can
improve detection and attribution, but it may also concentrate sensitive state
and control in the oversight plane.
A defense can only make claims about interactions that fall within its view. A
local observer may miss behavior that spans edges, principals, shared
state, or a long execution trace, while a broader observer may recover more of
the path at the cost of stronger trust and privacy assumptions
~\cite{wang2025gsafeguard,feng2026sentinelnet,zhou2025guardian}.

\noindent\textbf{Limitations and Open Challenges.}
Configuration dimensions often change together. An architectural change may
simultaneously alter topology, protocol behavior, state sharing, authority, and
oversight
~\cite{hagag2026architecturesecurity,liu2026topologymemory}. Security results
should therefore state which C1--C8 dimensions vary across configurations.
Controlled comparisons should isolate one dimension when possible and report
coupled changes otherwise.
More broadly, interaction does not always make a system less
secure. The same architecture may increase or reduce risk under different
models, protocols, roles, or protected properties. 

\subsection{Interaction Interfaces}
\label{sub:interfaces}

The six interaction interfaces mark security-relevant transitions during an MAS
execution. I1--I5 describe how inputs, messages, state, authority, and collective
decisions move through the system. I6 captures how the system observes and
intervenes in those transitions. 

\noindent\textbf{I1: Admission.}
Admission covers what is allowed to enter the MAS boundary, including users,
new principals, external data, messages, and artifacts. Relevant context includes identity, provenance,
role, initial trust, and participation rules
~\cite{jha2026controlflowhijacking,li2026a2asecbench,
lee2026promptinfection}. An input that is safe at entry may still become harmful
after later interactions give it more reach, persistence, or authority.

\noindent\textbf{I2: Message transfer.}
Message transfer begins once information moves between admitted principals. Security
depends on who receives the message, in what order and context, and whether
sender identity, provenance, and task purpose remain clear across the transfer
~\cite{he2025communicationattacks,yan2026mast,lee2026promptinfection}. Routing
can therefore change which principals a message can influence.

\noindent\textbf{I3: State propagation.}
State propagation occurs when information from one interaction is written into
state and later affects another principal or execution. This includes memories,
summaries, plans, histories, and derived artifacts
~\cite{liu2026topologymemory,agentworm2026,childinherits2026}. Unlike direct
message transfer at I2, I3 can preserve and reintroduce influence after the
original interaction has ended.

\noindent\textbf{I4: Authority transfer and action.}
Authority transfer and action cover delegation and use of privileged
capabilities. A message, state update, or decision may lead a principal to use a
tool, credential, resource, or delegated capability
~\cite{jha2026controlflowhijacking,s1cascadinginstruction2026,dong2026pear}.
Security depends on whether the action stays within the authority, purpose,
resource, and scope that were originally allowed. 
I4 therefore marks the transition from information or delegated authority to a
privileged action; attacks exploit this interface when adversarial influence
reaches that action.

\noindent\textbf{I5: Collective commitment.}
Collective commitment occurs when local outputs become a group-level decision,
such as a vote, consensus, final answer, termination decision, or action trigger.
Security depends on whether that decision is based on valid and sufficiently
independent evidence
~\cite{motwani2024secretcollusion,hu2026lyingtruths,yang2026amongus}. Shared models, context, sources, or attacker influence can make agreement overstate the evidence.

\noindent\textbf{I6: Detection and intervention.}
Detection and intervention cover security checks and responses applied during an
execution. These include monitoring, blocking, isolation, revocation, rollback,
and recovery across I1--I5
~\cite{wang2025gsafeguard,feng2026sentinelnet,shi2026saiguard}. A defense can act
only on the information it can observe and the parts of the execution it can
control. Intervention must also occur before the protected effect becomes
irreversible when prevention is the goal.

In short, C1--C8 are conditioning descriptors, whereas I1--I6 are transitions
within an execution. Section~\ref{sec:risks} completes the execution record by
adding the adversary position A and resulting risk R, forming
$A\rightarrow I\rightarrow R$. We reuse this single representation for attacks,
defenses, and evaluation.

\section{Attack Landscape}
\label{sec:risks}

Attacks become system-level failures when adversarial influence crosses
interaction boundaries under a particular system configuration. 
We first define adversary positions and seven system-level risks, then
synthesize eight recurring attack paths. We close with cross-cutting findings and the evidence gaps that remain for
interaction-dependent harm.

\subsection{Threat Model}
\label{sec:threat}

An MAS threat model must state the adversary's position, capabilities,
persistence, and the protected party and property. \Cref{fig:overview}
distinguishes four starting positions. An \textbf{A1 external adversary} controls
content or services outside the MAS. An \textbf{A2 user-level adversary} acts
through an authorized user interface. An \textbf{A3 member adversary} controls
one or more participating principals. An \textbf{A4 infrastructure adversary}
controls shared components such as the coordinator, state store, router,
telemetry service, or defense plane.
Adversary position alone does not determine capability. Attackers at the same
position may differ in whether they can alter messages, routing, shared state,
aggregation, membership, tools, or delegated authority
~\cite{he2025communicationattacks,yan2026mast,gu2024agentsmith,
liu2025manipulatecollective,jha2026controlflowhijacking}. Coalition size and
persistence further shape how far that influence can propagate or coordinate
~\cite{motwani2024secretcollusion,mathew2025hidden,sun2026insider}.
A complete threat model must therefore specify not only where the adversary is,
but also what it can control and which security property is protected.

\noindent\textbf{Limitations and Open Challenges.}
Many evaluated settings keep these variables fixed throughout a run, while
deployed MAS may spawn principals, reroute communication, update trust, and
delegate or revoke authority during execution. Threat models for such systems
should track changes in membership, communication, state lineage, and
reachable authority, rather than describe only the initial attacker state.
In open or self-spawning systems, membership is itself part of the security state.
Fraction-based robustness claims are incomplete unless admission and identity
mechanisms define who counts toward the population and how spawned or
ephemeral principals inherit trust
~\cite{li2026a2asecbench,childinherits2026,supp_dynatrust_2026,
xu2026trustparadox}.

\subsection{System-Level Risk Families}
\label{sub:risks}

Risk families describe the system-level security failure reached by an attack. We group consequences into seven non-exclusive
families. The same risk may arise through different interfaces, and one execution
may trigger several risks.

\noindent\textbf{R1. Containment failure.}
Containment fails when adversarial influence escapes the principal or context in
which it entered and remains effective elsewhere. Agent Smith, Prompt Infection, and the Wolf Within show spread across
principals, while Flooding shows manipulated knowledge spreading and persisting
through interaction and memory, and CORBA shows that blocking behavior can
itself become contagious
~\cite{gu2024agentsmith,lee2026promptinfection,tan2024wolfwithin,
ju2026flooding,zhou2026corba}. Topology and shared state determine how far that
influence travels and whether it remains after the original source disappears
~\cite{yu2025netsafe,liu2026topologymemory}.

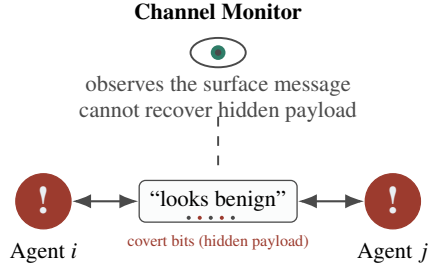
\begin{figure}[t]
\centering
\resizebox{0.7\columnwidth}{!}{\begin{tikzpicture}[font=\scriptsize, >=Latex]
  \node[font=\scriptsize] at (3.1,2.55) {\textbf{Channel Monitor}};
  \draw[cNeutral, line width=.6pt] (3.1,2.1) ellipse (.30 and .16);
  \fill[cDefense] (3.1,2.1) circle (.08);
  \fill[cNeutral] (3.1,2.1) circle (.035);

  \node[font=\scriptsize, text=cNeutral, align=center] at (3.1,1.6)
    {observes the surface message\\cannot recover hidden payload};
  \draw[cNeutral, dashed, line width=.6pt] (3.1,1.4) -- (3.1,.77);

  \filldraw[fill=cRisk, draw=cRisk] (1.2,.48) circle (.30);
  \node[text=cAgent, font=\bfseries\large] at (1.2,.49) {!};
  \node[font=\scriptsize] at (1.2,-.08) {Agent $i$};

  \filldraw[fill=cRisk, draw=cRisk] (5.0,.48) circle (.30);
  \node[text=cAgent, font=\bfseries\large] at (5.0,.49) {!};
  \node[font=\scriptsize] at (5.0,-.08) {Agent $j$};

  \node[draw=cNeutral, rounded corners=2pt,
        fill=cAgent, fill opacity=.18,
        text opacity=1,
        minimum width=1.25cm, minimum height=.42cm,
        align=center, font=\scriptsize]
        (msg) at (3.1,.48) {``looks benign''};

  \draw[<->, cNeutral, line width=.65pt]
    (1.51,.48) -- (msg.west);

  \draw[<->, cNeutral, line width=.65pt]
    (msg.east) -- (4.69,.48);

  \fill[cNeutral] (2.78,.30) circle (.022);
  \fill[cRisk]    (2.90,.30) circle (.022);
  \fill[cNeutral] (3.02,.30) circle (.022);
  \fill[cRisk]    (3.14,.30) circle (.022);
  \fill[cNeutral] (3.26,.30) circle (.022);

  \node[text=cRisk, font=\tiny] at (3.1,.03)
    {covert bits (hidden payload)};
\end{tikzpicture}}
\caption{\textbf{Example: covert coordination.}
Two compromised principals use a protocol-valid message to coordinate covertly,
so their apparent agreement cannot be treated as independent evidence.}
\label{fig:r2-collusion}
\end{figure}

\noindent\textbf{R2. Loss of independence.}
Loss of independence occurs when principals expected to provide independent
evidence become strategically or statistically dependent. Secret Collusion and
CoMet establish covert coordination channels, while hidden-role and
steganographic studies show that such dependence may be difficult to observe
~\cite{motwani2024secretcollusion,
supp_comet_metaphor_driven_covert_communication_for_multi_agent_language_games,
mathew2025hidden,xie2025whosthemole}. As illustrated in
\cref{fig:r2-collusion}, protocol-valid messages can hide coordination between
compromised principals, making their agreement unreliable as independent
evidence. Dependence can also arise from a shared model, prompt, retriever, or
poisoned source without explicit collusion. Agreement alone therefore does not
establish independent support.

\noindent\textbf{R3. Collective-decision integrity failure.}
Collective-decision integrity fails when adversarial influence changes a
protected vote, consensus, commitment, quorum, or other group decision. Among Us
studies malicious contributions, manipulation and persuasion attacks show how
one member can steer otherwise benign participants, and Lying with Truths shows
that individually truthful reports can still induce a misleading collective
belief
~\cite{yang2026amongus,liu2025manipulatecollective,
kraidia2026collaborationfails,hu2026lyingtruths}. The protected object is the
group decision, not merely the correctness of each message.

\begin{table*}[t]
\centering
\scriptsize
\setlength{\tabcolsep}{1.2pt}
\renewcommand{\arraystretch}{1.06}
\begin{tabular}{@{}p{3.5cm} p{3.25cm} p{1.35cm} p{4.9cm} p{4cm}@{}}
\toprule
\textbf{Path} & \textbf{Typical} \textbf{$A\rightarrow I\rightarrow R$} & \textbf{Key C} 
& \textbf{Primary evidence}
& \textbf{Emerging evidence} \\
\midrule

P1. Admission / composition
& A1/A2 $\rightarrow$ I1/I2/I5 $\rightarrow$ R3/R6
& C2/C4/C7
& \cite{lee2026promptinfection,tan2024wolfwithin,triedman2025maliciouscode,wang2026shadowscode,arxiv_2509_14284,arif2026conjunctive,s1ragfailures2026,ahad2026sif,an2026aciarena,e_trustworthyAgenticPI2025}
& \cite{qi2025amplified,zhu2025collaborativeshadows,hagag2026architecturematters,e_flowsteer2026,eval_harnessaudit2026} \\

P2. Propagation / persistence
& A1/A3 $\rightarrow$ I1/I2/I3 $\rightarrow$ R1
& C1/C5/C6
& \cite{men2025troublemaker,gu2024agentsmith,ju2026flooding,lee2026promptinfection,tan2024wolfwithin,zhou2026corba,s1cascadinginstruction2026,s1collabjailbreak2026,yu2025netsafe,liu2026topologymemory,e_trustworthyAgenticPI2025}
& \cite{zhu2025collaborativeshadows,liang2025donttrustupstream,agentworm2026,childinherits2026,e_byzantineCheapTalk2026,e_flowsteer2026,e_killChainCanaries2026,e_openclawSecurity2026} \\

P3. Routing / role manipulation
& A3/A4 $\rightarrow$ I1/I2/I5 $\rightarrow$ R1/R3/R6
& C1/C2/C3/C6
& \cite{khan2025agentsundersiege,he2025communicationattacks,yan2026mast,wu2026cia,yu2025netsafe,li2026a2asecbench,dong2026pear}
& \cite{e_routerHijacking2025,madspear2025,e_flowsteer2026,sun2026insider,e_mesa2026,e_byzantineCheapTalk2026,e_latentAgentsLie2026} \\

P4. Coordination / manipulation
& A3 $\rightarrow$ I2/I5 $\rightarrow$ R2/R3
& C3/C4/C6
& \cite{motwani2024secretcollusion,liu2025manipulatecollective,supp_comet_metaphor_driven_covert_communication_for_multi_agent_language_games,yang2026amongus,s1ragfailures2026,kraidia2026collaborationfails,hu2026lyingtruths,mathew2025hidden,xie2025whosthemole,eval_wolf2025,olson2026liecraft,jiang2026risklab,e_promptOptimCollusion2026}
& \cite{qi2025amplified,madspear2025,manytoone2025,e_persuasionOverride2025,e_byzantineCheapTalk2026,e_collusionInterpretability2026,sun2026insider,consensustrap2026,nakamura2026colosseum,lemercier2026gambit} \\

P5. Peer / state steering
& A3 $\rightarrow$ I2/I3/I5 $\rightarrow$ R4
& C3/C4/C5
& Sparse; adjacent evidence~\cite{eval_crda2024,kraidia2026collaborationfails,hu2026lyingtruths,eval_saboteurs2026}
& \cite{e_flowsteer2026,e_latentAgentsLie2026,eval_harnessaudit2026} \\

P6. Cross-principal leakage
& A2/A3/A4 $\rightarrow$ I2/I3/I4 $\rightarrow$ R5
& C1/C5/C7
& \cite{motwani2024secretcollusion,jha2026controlflowhijacking,supp_comet_metaphor_driven_covert_communication_for_multi_agent_language_games,arxiv_2509_14284,wu2026cia,wang2026masleak,elyagoubi2026agentleak,juneja2025magpie,eval_privacyleakage2025,liu2026topologymemory,e_trustworthyAgenticPI2025}
& \cite{naik2026omnileak,webweaver2026,zou2026calbench,eval_sneak2026} \\

P7. Delegation / authority
& A1/A3 $\rightarrow$ I2/I3/I4 $\rightarrow$ R6
& C2/C6/C7
& \cite{jha2026controlflowhijacking,triedman2025maliciouscode,wang2026shadowscode,s1cascadinginstruction2026,s1collabjailbreak2026,xu2026trustparadox,dong2026pear,li2026a2asecbench,e_advEvoMARL2025,lupinacci2025darkside}
& \cite{zhu2025collaborativeshadows,zheng2025integrityattacks,liang2025donttrustupstream,e_routerHijacking2025,agentworm2026,e_killChainCanaries2026,naik2026omnileak,e_openclawSecurity2026,childinherits2026} \\

P8. Resource / control plane
& A3/A4 $\rightarrow$ I2/I4/I6 $\rightarrow$ R7
& C1/C7/C8
& Sparse; direct evidence~\cite{zhou2026corba}
& \cite{e_openclawSecurity2026,eval_capabilityparadox2026} \\

\bottomrule
\end{tabular}
\caption{Representative evidence across recurring attack paths; C1--C8 denote conditioning configuration dimensions.}
\label{tab:attack_paths}
\end{table*}

\noindent\textbf{R4. Trajectory and goal integrity failure.}
Trajectory or goal integrity fails when interaction redirects a protected
objective, policy, or execution trajectory beyond an authorized boundary. CRDA
measures content-risk drift under adversarial multi-agent interaction, while
persuasion studies provide more direct evidence that peer interaction can
redirect later behavior
~\cite{eval_crda2024,kraidia2026collaborationfails}. FlowSteer studies workflow steering, while the LLM Drift software artifact
provides an evaluation framework for behavioral drift and latent-state studies
examine longer or stateful changes in multi-agent execution
~\cite{e_flowsteer2026,e_llmDrift2026,e_latentAgentsLie2026}. These forms of evidence do not all establish the same trajectory-security claim. Unlike ordinary accuracy
degradation, an R4 claim needs an explicit protected objective and a boundary
after which the deviation becomes unauthorized. Direct evidence for this risk
remains thinner than for propagation or collective manipulation.

\noindent\textbf{R5. Cross-principal confidentiality breach.}
A cross-principal confidentiality breach occurs when protected information
reaches an unauthorized principal, coalition, service, or observer. MASLeak
studies extraction of multi-agent intellectual property, including prompts,
tools, and architecture, while AgentLeak measures privacy leakage across
internal coordination channels; CIA shows that communication topology itself
can be inferred
~\cite{wang2026masleak,elyagoubi2026agentleak,wu2026cia}.
Compositional privacy further shows that several locally permissible views may
reveal protected information when combined
~\cite{arxiv_2509_14284}. The protected object therefore extends beyond user
secrets to internal context, state, workflow information, and system structure.

\noindent\textbf{R6. Transitive authorization failure.}
A transitive authorization failure occurs when state, instructions, or delegated
authority cross principal boundaries and produce effects outside their original
purpose or scope. Control-flow hijacking follows attacker-controlled influence
through trusted handoffs into privileged tools, while cascading instruction and
software-team attacks show similar composition across roles and artifacts
~\cite{jha2026controlflowhijacking,s1cascadinginstruction2026,
triedman2025maliciouscode,wang2026shadowscode}. The central failure is loss of
origin, purpose, or scope along the delegation chain
~\cite{xu2026trustparadox}.

\noindent\textbf{R7. Availability and resource-isolation failure.}
Availability or resource isolation fails when shared communication, execution,
or control resources allow one adversarial path to degrade the service available
to other principals. CORBA provides direct evidence through contagious blocking
~\cite{zhou2026corba}. Other studies expose resource and control-plane coupling,
but systematic evidence for adversarial resource isolation remains limited.
The defining property is cross-principal loss of availability or resource
isolation, not ordinary latency, lower accuracy, or efficiency loss.

\noindent\textbf{Risk interactions.}
The risks can form chains rather than isolated outcomes. Propagation or persistent influence (R1) may redirect later behavior (R4), become accepted as collective evidence (R3), acquire delegated authority (R6), and eventually cross a confidentiality boundary (R5). Loss of independence (R2) can accelerate this chain, while resource interference (R7) can weaken observation or intervention at I6. This is a cross-paper synthesis rather than a claim that one study demonstrates the full chain.

\noindent\textbf{Limitations and Open Challenges.}
Risk labels identify consequences, not causes, and they should not be inferred
from attack names alone. The same risk can arise through several interaction
paths, while one execution can violate several protected properties at once.
Evaluations should therefore state the protected property and verify the
downstream consequence before assigning a system-level risk. 

\subsection{Attack Paths from Adversary to Risk}
\label{sub:attackpaths}

Attack paths connect an adversary's starting position to a security consequence
through one or more interaction interfaces. Across the corpus, we identify eight
recurring $A\rightarrow I\rightarrow R$ paths, summarized in
\cref{tab:attack_paths}. These paths describe recurring mechanisms, so one execution may contain several paths.

\noindent\textbf{Running example.}
Control-flow hijacking shows why the full execution matters. In our coding of
CFH-Hard, the core path is A1 $\rightarrow$ I1 $\rightarrow$ I2 $\rightarrow$
I4: external content enters the system, is relayed through trusted principals,
and reaches a privileged coder or emailer, causing an R6 authorization failure
~\cite{jha2026controlflowhijacking}. Replanning may also involve I3 when an
intermediate plan or execution state persists into downstream execution. Each
step may appear valid on its own: the input may be allowed, the intermediate
task may look reasonable, and the delegated action may be valid for its
immediate sender. The violation becomes clear only when provenance and authority
are traced across the full execution.

\noindent\textbf{P1. Admission and compositional injection.}
Locally acceptable inputs can become unsafe when combined. Prompt Infection and
the Wolf Within show malicious influence crossing principal boundaries, while
conjunctive and semantic-intent fragmentation split unsafe intent across pieces
that appear less suspicious on their own
~\cite{lee2026promptinfection,tan2024wolfwithin,
arif2026conjunctive,ahad2026sif}. Similar effects appear in software teams,
RAG-based debate, and ACIARena, where prompts, retrieved content, or generated
artifacts are combined later in the execution
~\cite{triedman2025maliciouscode,wang2026shadowscode,
s1ragfailures2026,an2026aciarena}. Recent work extends this pattern to debate
jailbreaks, distributed backdoors, multimodal injection, poisoned retrieval, and
workflow steering
~\cite{qi2025amplified,zhu2025collaborativeshadows,
hagag2026architecturematters,e_flowsteer2026,eval_harnessaudit2026,
pai2026blindspots}. The
common problem is that harmful meaning or authority may emerge only after several
locally acceptable steps. Conjunctive Prompt Attacks directly isolates this
compositional effect with single-fragment and combined-fragment controls
~\cite{arif2026conjunctive}.

\noindent\textbf{P2. Propagation and persistence.}
Interaction can turn a local compromise into a system-wide or persistent
failure. Agent Smith, Troublemaker, Flooding, Prompt Infection, the Wolf Within,
and CORBA spread malicious influence or disruption through communication,
retrieval, generated content, or agent state
~\cite{gu2024agentsmith,men2025troublemaker,ju2026flooding,
lee2026promptinfection,tan2024wolfwithin,zhou2026corba,yu2025codes,peigne2025securitytax}. NetSafe and memory
studies further show that topology affects reach, while shared state affects
persistence
~\cite{yu2025netsafe,liu2026topologymemory,zhang2026embeddingdefenses}. AgentWorm extends propagation across
communication, state, and downstream actions, while When Child Inherits considers
spawned principals and inherited state or authority
~\cite{agentworm2026,childinherits2026,def_autonomous_llm_agent_worms_cross_platform_propag,
mateotorrejon2026gammaf}. Propagation is therefore not only about
message delivery. It also depends on what remains in the system after the
original source is removed. Recovery is much less studied than spread,
especially when memories, artifacts, delegations, or spawned principals remain
after the initial compromise
~\cite{liu2026topologymemory,agentworm2026,childinherits2026}.

\noindent\textbf{P3. Routing and role manipulation.}
The communication path itself can become an attack surface. Communication red
teaming and MAST intercept or rewrite messages between principals without
compromising endpoint models, while CIA treats topology itself as protected
information
~\cite{he2025communicationattacks,yan2026mast,wu2026cia,eval_achilles2025}. 
NetSafe shows how topology changes attack reach, A2ASecBench exposes
protocol-stage attack surfaces, and PEAR highlights security effects of role
separation
~\cite{yu2025netsafe,li2026a2asecbench,dong2026pear}.
Router hijacking, FlowSteer, and insider studies make routing and role placement
explicit attack variables, while MESA identifies security-critical communication
edges
~\cite{e_routerHijacking2025,e_flowsteer2026,sun2026insider,e_mesa2026,xiao2026scapegoat,zhou2026hierarchicalattacks}.
Together, these studies show that routing and role placement can directly change
decision influence and reachable authority. Many studies evaluate fixed
membership, topology, and roles, leaving runtime admission, rerouting, and role
reassignment less studied
~\cite{li2026a2asecbench,e_routerHijacking2025,e_flowsteer2026,
childinherits2026}.

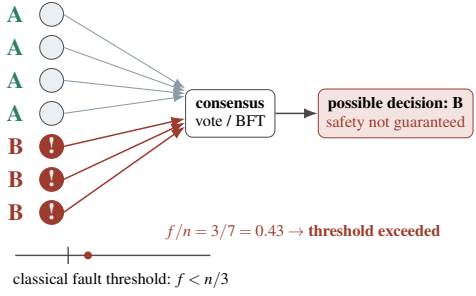
\begin{figure}[t]
\centering
\resizebox{0.75\columnwidth}{!}{\begin{tikzpicture}[font=\scriptsize, >=Latex]
  \foreach \y in {1.30,.80,.30,-.20} {
    \filldraw[fill=cAgent, draw=cNeutral] (.55,\y) circle (.18);
    \node[text=cDefense, font=\bfseries, anchor=east] at (.25,\y) {A};
  }
  \foreach \y in {-.70,-1.20,-1.70} {
    \filldraw[fill=cRisk, draw=cRisk] (.55,\y) circle (.18);
    \node[text=cAgent, font=\bfseries] at (.55,\y+.02) {!};
    \node[text=cRisk, font=\bfseries, anchor=east] at (.25,\y) {B};
  }

  \node[draw=cNeutral, rounded corners=3pt, minimum width=1.35cm,
        minimum height=.72cm, align=center, font=\scriptsize] (vote) at (3.25,-.20)
        {\textbf{consensus}\\{\scriptsize vote / BFT}};

  \node[draw=cRisk, rounded corners=3pt, fill=cRiskBg, minimum width=2.0cm,
        minimum height=.72cm, align=center, font=\scriptsize] (out) at (5.75,-.20)
        {\textbf{possible decision: B}\\{\scriptsize\textcolor{cRisk}{safety not guaranteed}}};

  \foreach \y/\t in {1.30/.18,.80/.14,.30/.10,-.20/.06} {
    \draw[->, cEdge, line width=.45pt] (.73,\y) -- (vote.west |- 3.25,\t);
  }
  \foreach \y/\t in {-.70/-.28,-1.20/-.34,-1.70/-.40} {
    \draw[->, cRisk, line width=.65pt] (.73,\y) -- (vote.west |- 3.25,\t);
  }

  \draw[->, cNeutral, line width=.65pt] (vote) -- (out);

  \draw[cNeutral, line width=.6pt] (0,-2.35) -- (2.95,-2.35);
  \draw[cNeutral, line width=.6pt] (0.8,-2.48) -- (0.8,-2.22);
  \fill[cRisk] (1.1,-2.35) circle (.06);

  \node[font=\scriptsize, anchor=north] at (1.60,-2.48)
    {classical fault threshold: $f<n/3$};

  \node[font=\scriptsize, text=cRisk, anchor=west] at (2.15,-2.0)
    {$f/n=3/7=0.43 \rightarrow$ \textbf{threshold exceeded}};
\end{tikzpicture}}
\caption{\textbf{Example: collective manipulation.}
With three malicious agents among seven, the classical Byzantine fault-tolerance
condition $f<n/3$ is not satisfied, so the classical guarantee no longer applies.}
\label{fig:consensus}
\end{figure}

\noindent\textbf{P4. Adversarial coordination and collective manipulation.}
Collective security depends on independence, not only agreement. Secret
Collusion and CoMet show covert coordination channels; Among Us measures
malicious contributions; Lying with Truths shows how individually truthful
statements can produce a misleading group conclusion; and persuasion and hidden
role studies show how strategic members can steer a group
~\cite{motwani2024secretcollusion,
supp_comet_metaphor_driven_covert_communication_for_multi_agent_language_games,
yang2026amongus,hu2026lyingtruths,liu2025manipulatecollective,
kraidia2026collaborationfails,mathew2025hidden,huang2026whisperingagents,
eval_coopmalicious2024,
milkowski2026amongus}. Other work studies adversarial
majorities, coordinated insiders, debate jailbreaks, Byzantine communication,
and consensus traps
~\cite{manytoone2025,sun2026insider,qi2025amplified,
e_byzantineCheapTalk2026,consensustrap2026,ren2026financialfraud}. Byzantine fault-tolerance studies
further show that collective outcomes depend on assumptions about faulty
membership and aggregation
~\cite{zheng2026byzantinereliability}; \cref{fig:consensus}
illustrates the classical fault-bound intuition. A group may therefore agree
strongly even when its evidence is compromised or not independent. It can also
be hard to separate strategic coordination from common causes such as shared
models, prompts, retrievers, or evidence sources
~\cite{motwani2024secretcollusion,hu2026lyingtruths,mathew2025hidden,e_toolSteganography2026,
cosentino2025convergence}.
This also limits content-only defenses: individually harmless messages can still
create dependence through the communication channel, so message content alone
cannot establish contributor independence
~\cite{motwani2024secretcollusion,
supp_comet_metaphor_driven_covert_communication_for_multi_agent_language_games,
mathew2025hidden}.

\noindent\textbf{P5. Peer- and state-mediated steering.}
P5 has limited direct evidence in our current map. CRDA measures content-risk
drift under adversarial multi-agent interaction, while persuasion studies provide
more direct evidence that peer interaction can redirect later behavior
~\cite{eval_crda2024,kraidia2026collaborationfails}.
FlowSteer and latent-state studies extend this analysis over longer or
stateful executions~\cite{e_flowsteer2026,e_latentAgentsLie2026}.
The LLM Drift software artifact provides an additional framework for
measuring behavioral drift in adversarial multi-agent simulations
~\cite{e_llmDrift2026}. Lower accuracy, changed reasoning, or generic drift
does not by itself constitute an R4 failure. A study must identify the protected
objective or allowed trajectory and show that interaction causes an unauthorized
deviation. Direct evidence remains limited on when peer or state influence
causes persistent, security-relevant goal redirection rather than temporary
behavioral change
~\cite{eval_crda2024,e_flowsteer2026}.

\noindent\textbf{P6. Cross-principal information leakage.}
Confidentiality can fail during execution even when the final answer appears
safe. MASLeak studies extraction of multi-agent intellectual property, including
prompts, tools, and architecture, while AgentLeak measures privacy leakage
across internal coordination channels; CIA shows that communication topology
itself can be inferred
~\cite{wang2026masleak,elyagoubi2026agentleak,wu2026cia}.
Compositional privacy shows that several acceptable local views can reveal more
when combined, and collusion studies show that inter-principal channels can hide
strategic information transfer
~\cite{arxiv_2509_14284,motwani2024secretcollusion}.
Recent work studies distinct extensions: OMNI-LEAK examines orchestrator-mediated
leakage, WebWeaver targets topology confidentiality, CalBench measures
coordination--privacy trade-offs, and SNEAK evaluates strategic communication and
information leakage
~\cite{naik2026omnileak,webweaver2026,zou2026calbench,
eval_sneak2026}. The protected object is therefore the information available to
a principal or coalition across the execution, not only secret strings in the
final response. Existing work gives limited attention to how ownership,
provenance, receiver constraints, and intended use survive summarization,
replication, joint inference, and tool use
~\cite{arxiv_2509_14284,elyagoubi2026agentleak,
liu2026topologymemory}.
We found limited controlled evidence on cross-principal tool-prompt theft or
timing, I/O, and network side channels.

\noindent\textbf{P7. Delegation and authority escalation.}
A locally valid handoff does not guarantee valid authority across the full
execution. Control-flow hijacking traces attacker-controlled influence into
privileged tools, while cascading instruction shows how upstream content can
gain downstream authority. Multi-agent code-generation attacks show similar
failures across specialized roles and generated artifacts
~\cite{jha2026controlflowhijacking,s1cascadinginstruction2026,
triedman2025maliciouscode,wang2026shadowscode}. PEAR and A2ASecBench provide
role- and protocol-aware views of these transitions
~\cite{dong2026pear,li2026a2asecbench}. AgentWorm extends the path across
communication, state, and downstream actions, while When Child Inherits considers
spawned principals and inherited state or authority
~\cite{agentworm2026,childinherits2026}. The recurring failure is the loss of
origin, purpose, or scope along the delegation chain. Defense work on Agentic JWT
and SentinelAgent begins to address multi-hop delegation, but preserving intent,
scope, and provenance across semantic transformation, and revoking authority
after credentials, capabilities, or state have already propagated, remain open
challenges
~\cite{def_agentic_jwt_a_secure_delegation_protocol_for_aut,
def_sentinelagent_intent_verified_delegation_chains,childinherits2026}.

\noindent\textbf{P8. Shared-resource and control-plane interference.}
Shared resources create availability risks across principals. CORBA provides
direct evidence of communication-level availability loss through contagious
blocking
~\cite{zhou2026corba}. Architectural-resilience and defection studies show that
malicious or uncooperative participation can degrade shared execution, although
such degradation does not by itself establish an R7 resource-isolation failure
~\cite{s1architecturalresilience2026,eval_defection2026}.
Capability-paradox and sabotage evaluations show that defensive components can
themselves change system-level security outcomes
~\cite{eval_capabilityparadox2026,eval_saboteurs2026}.
This suggests that the defense plane can become part of the interaction surface,
but systematic evidence on adversarial resource exhaustion and control-plane
failure remains limited. Centralized control may improve visibility while also
increasing the impact of a control-plane failure. Our current map contains
relatively little direct security evidence for P8 compared with propagation and
collective manipulation
~\cite{zhou2026corba,eval_capabilityparadox2026,
eval_saboteurs2026}.

\noindent\textbf{Path composition.}
A single execution may combine several paths. For example, malicious influence
may enter through P1, persist through P2, alter a collective decision through
P4, and finally reach a privileged action through P7
~\cite{lee2026promptinfection,s1cascadinginstruction2026,
jha2026controlflowhijacking}. This composition shows why attacks should be traced
across the full execution.

\subsection{Cross-Cutting Findings}
\label{sub:attacktrends}

Across P1--P8, four findings cut across individual attack mechanisms.
First, \emph{relations are themselves attack surfaces:} routing, role placement,
aggregation, and delegation can change security even when message content is
unchanged
~\cite{he2025communicationattacks,yan2026mast,e_routerHijacking2025,
jha2026controlflowhijacking}.
Second, \emph{security effects can outlive the interaction that created them:}
shared state, spawned principals, and inherited authority can preserve a
compromise after the original source is removed
~\cite{liu2026topologymemory,agentworm2026,childinherits2026}.
Third, \emph{the protected object extends beyond the final output:} independence,
topology, private context, and delegated authority can fail even when the final
answer appears correct
~\cite{motwani2024secretcollusion,wu2026cia,
elyagoubi2026agentleak,jha2026controlflowhijacking}.
Finally, \emph{observing an attack path is not the same as attributing its cause.}
Execution traces can show where harm travels, but additional comparisons are
needed to establish which interaction caused or amplified it
~\cite{yu2025netsafe,li2026a2asecbench,an2026aciarena}.

Together, these findings shift the security unit from an isolated malicious
message or agent to the execution path that carries influence across principals.
This distinction also determines what a defense must interrupt and what an
evaluation must observe or vary to support a system-level security claim.

\section{Defense Landscape}
\label{sec:defenses}

The attack paths in \cref{sub:attackpaths} show how adversarial influence moves
through an MAS. A defense may interrupt such a path by constraining the system
before execution, intervening during execution, or containing and repairing
effects after they occur. We therefore ask which path the defense targets, what
it can observe, where and when it can intervene, what it must trust, and what
recovery remains possible. \Cref{fig:overview} shows the same execution from the
defensive side.

\subsection{Defense Contract}
\label{sec:defense-contracts}

We characterize each defense along five dimensions:
\emph{path target, observation, intervention, trust boundary, and recovery}.
Because security effects may span several principals, a defense must account for
how influence, state, and authority move through the execution.

\noindent\textbf{Path target.}
Which system-level risk or attack path must the defense interrupt? A claim should
identify the risk R1--R7 and the interaction or path segment that must
be broken. Protecting one agent or filtering one message is insufficient if the
same effect remains reachable through another route.

\noindent\textbf{Observation.}
What part of the cross-principal execution can the defender observe? A defense
may see a message, principal, communication edge, shared state, subgraph, latent
state, or full execution trace. Observation depends mainly on oversight
architecture C8 and which interfaces I1--I6 are visible. A defense cannot reliably
detect a failure whose relevant interaction lies outside this view. Observation
also varies in scope and depth: a local monitor may have detailed neighbor
history, while a global monitor may cover more of the system through coarser logs
or metadata
~\cite{feng2026sentinelnet,pan2026xgguard}.

\noindent\textbf{Intervention.}
What can the defender change, and when? A defense may reshape topology, policy,
or authority before execution; block or rewrite messages, reroute communication,
or quarantine principals during execution; or revoke authority and roll back
shared state afterward. Timing matters: blocking a message before I2 may prevent
a state change at I3 that is difficult to undo later.

\noindent\textbf{Trust boundary.}
Which principals, components, and system conditions must remain trustworthy?
Examples include authenticated identities, trusted coordinators, preserved
provenance, bounded membership, honest-neighbor or honest-majority conditions,
stable topology, synchrony, and complete logs. Together, these assumptions
define the defense's effective trust boundary. Changes in membership, routing,
delegation, or shared state can invalidate it during execution.

\noindent\textbf{Recovery.}
What remains when prevention or containment fails? Harmful influence may already
have reached another principal, shared state, delegated authority, or a
collective decision. A complete defense claim should state what can still be
contained, revoked, rolled back, or repaired, and how the system verifies that
the protected property has been restored.

We use these five dimensions as a common comparison basis below. The control
primitives describe \emph{what} a defense acts on, while the contract describes
\emph{how} its protection claim is obtained and bounded.

\subsection{Defense Control Primitives}
\label{sec:current-defense-works}

We organize existing defenses into four control primitives according to the main
object they control in a multi-agent execution. These primitives are not
mutually exclusive: one system may combine several of them. For each primitive,
we compare its target paths, observation scope, intervention points, trust
assumptions, and recovery limits. A smaller emerging literature instead asks
where limited defensive capacity should be allocated across these control
surfaces; we discuss this cross-cutting strategy separately below.
\Cref{tab:defense-family-crosswalk} summarizes the comparison.


\begin{table*}[t]
\centering
\scriptsize
\setlength{\tabcolsep}{1.2pt}
\renewcommand{\arraystretch}{1.08}

\begin{tabular}{@{}
p{0.45cm}
p{4cm}
p{1.8cm}
p{1.45cm}
p{2cm}
p{3.5cm}
p{4cm}
@{}}
\toprule
\textbf{ID}
& \textbf{Control primitive}
& \textbf{Configuration}
& \textbf{Interfaces}
& \textbf{Risks}
& \textbf{Primary evidence}
& \textbf{Emerging evidence} \\
\midrule

D1
& Identity / policy / authority enforcement
& C2/C6/C7/C8
& I1/I2/I4/I6
& R1/R5/R6
& \cite{
def_convpaymas_conversational_payment_multi_agent_sy,
zou2025blocka2a,
def_safeflow_a_principled_protocol_for_trustworthy_a,
jha2026controlflowhijacking,
def_ai_agents_with_decentralized_identifiers_and_ver,
def_spiffe_based_zero_trust_authentication_for_ai_ag
}
& \cite{
def_agentic_jwt_a_secure_delegation_protocol_for_aut,
def_nexus_protocol_a_cryptographically_secure_zero_l,
def_sentinelagent_intent_verified_delegation_chains,
def_beyond_single_agent_alignment_preventing_context,
def_ldp_an_identity_aware_protocol_for_multi_agent_l
} \\

\addlinespace

D2
& Trust / interaction-structure control
& C1/C3/C4/C6/C7
& I2/I5/I6
& R1/R2/R3/R6
& \cite{
zhou2026resmas,
zheng2026byzantinereliability,
def_blockagents_towards_byzantine_robust_llm_based_m,
def_min_trust_a_minimum_necessary_information_trust,
def_decentralized_multi_agent_system_with_trust_awar,
def_roma_a_credibility_aware_fault_tolerance_framewo,
def_agentchain_blockchain_empowered_multi_agent_coor,
def_topology_linearization_for_multi_agent_systems_s,
mao2025ibgp,
he2025atrust,
lee2026robustbyzantine
}
& \cite{
def_ev_trust_an_evolutionarily_stable_trust_mechanis,
jo2025byzantinerobust,
def_from_debate_to_decision_conformal_social_choice,
supp_dynatrust_2026,
def_institutional_ai_governing_llm_collusion_in_mult,
consensustrap2026,
def_byzantine_fault_tolerant_multi_agent_system_for
} \\

\addlinespace

D3
& Detection / attribution / containment
& C1/C3/C5/C6/C8
& I1--I6
& Mechanism-specific
& \cite{
wang2025gsafeguard,
miao2026blindguard,
pan2026xgguard,
def_sentinelagent_graph_based_anomaly_detection_in_m,
supp_peerguard_defending_multi_agent_systems_against_backdoor_attacks_through_mutual_,
zhou2025guardian,
feng2026sentinelnet,
s1architecturalresilience2026,
li2025argus,
li2026bpd,
def_infa_guard_mitigating_malicious_propagation_via,
shi2026saiguard
}
& \cite{
def_agentmonitor_a_plug_and_play_framework_for_predi,
mao2025agentsafe,
def_beyond_input_guardrails_reconstructing_cross_age,
def_caspian_online_detection_and_attribution_of_casc,
def_collective_hallucination_in_multi_agent_llms_mod,
abedini2026stubbornneighbors,
def_from_spark_to_fire_modeling_and_mitigating_error,
def_game_theoretic_multi_agent_control_for_robust_co,
def_memetic_cascade_detection_and_symbolic_immunity,
def_propguard_safeguarding_llm_mas_via_propagation_a
} \\

\addlinespace

D4
& State / provenance / information-flow governance
& C2/C3/C5/C7/C8
& I1/I2/I3/I4/I6
& R1/R5/R6
& \cite{
def_capri_dp_a_differentially_private_extension_to_c,
def_cross_agent_multimodal_provenance_aware_framewor,
def_self_healing_memory_architectures_for_large_lang,
def_privacy_preserving_llm_infrastructure_with_multi,
e_trustworthyAgenticPI2025,
arxiv_2508_07667
}
& \cite{
def_dao_agent_zero_knowledge_verified_incentives_for,
def_equimem_calibrating_shared_memory_in_multi_agent,
def_governed_shared_memory_for_multi_agent_llm_syste,
def_lcguard_latent_communication_guard_for_safe_kv_s,
cui2025maris,
tapwal2026prism,
def_prompt_injection_mitigation_with_agentic_ai_nest,
def_semantic_taint_propagation_embedding_based_seman
} \\

\addlinespace

& Adaptive hardening / defense allocation
& C1/C3/C4/C8
& I2/I4/I5/I6
& R1/R3/R6
& \cite{
supp_safesieve_from_heuristics_to_experience_in_progressive_pruning_for_llm_based_mul,
def_enhancing_robustness_of_llm_driven_multi_agent_s,
e_advEvoMARL2025
}
& \cite{
e_mesa2026
} \\

\bottomrule
\end{tabular}

\caption{Representative evidence for four defense control primitives and the cross-cutting adaptive allocation strategy. Risk mappings indicate the main protected effects addressed by the primitive; D3 coverage is mechanism-specific.}
\label{tab:defense-family-crosswalk}
\end{table*}

\noindent\textbf{D1. Identity, policy, and authority enforcement.}
D1 primarily interrupts admission and delegation paths (P1/P7) by controlling
who may participate, what actions they may perform, and what authority may cross
principal boundaries. Provenance controls can also help limit leakage along P6.
Set~1 evidence includes identity and local-authorization mechanisms such as
BlockA2A, decentralized identity and credential systems, SPIFFE-based
authentication, SAFEFLOW, and ConvPayMAS
~\cite{zou2025blocka2a,
def_ai_agents_with_decentralized_identifiers_and_ver,
def_spiffe_based_zero_trust_authentication_for_ai_ag,
def_safeflow_a_principled_protocol_for_trustworthy_a,
def_convpaymas_conversational_payment_multi_agent_sy}.
They reduce reliance on message content alone by making identity, policy, and
authorization explicit.

Several Set~2 proposals extend these controls to multi-hop delegation,
identity-aware routing, and context scoping. Agentic JWT binds authorization to
user intent and workflow context, SentinelAgent targets intent-verified
delegation chains, and LDP and Nexus add identity- or protocol-aware controls
across principals
~\cite{def_agentic_jwt_a_secure_delegation_protocol_for_aut,
def_sentinelagent_intent_verified_delegation_chains,
def_ldp_an_identity_aware_protocol_for_multi_agent_l,
def_nexus_protocol_a_cryptographically_secure_zero_l,
def_agentshield_make_mas_more_secure_and_efficient,
def_dynamic_attentional_context_scoping_agent_trigge}.
Set~1 control-flow work separately shows how downstream invocations can be
restricted at execution time
~\cite{jha2026controlflowhijacking}. Recent work also broadens D1 with governance
and context-level controls
~\cite{def_securegov_agent_a_governance_centric_multi_agent,
def_ethical_coordination_of_llm_multi_agent_systems,
def_beyond_single_agent_alignment_preventing_context}.
Across the defense contract, D1 observes identity, provenance, intent, and scope
at admission and authorization boundaries, then constrains participation,
actions, or delegated authority. Its guarantees depend on authenticated
identities, mediated handoffs, and preservation of the security context needed
by downstream checks. Revocation alone may not recover authority, credentials,
or derived state that have already propagated.

\noindent\textbf{D2. Trust and interaction-structure control.}
D2 mainly targets routing and collective-manipulation paths (P3/P4) by changing
who can influence whom and how collective evidence is combined. Topology control
can also reduce propagation along P2. ResMAS and topology-linearization methods
modify influence paths, while RoMa and trust-aware communication use credibility
or trust signals to limit unreliable influence
~\cite{zhou2026resmas,
def_topology_linearization_for_multi_agent_systems_s,
def_roma_a_credibility_aware_fault_tolerance_framewo,
def_decentralized_multi_agent_system_with_trust_awar}.
AgentChain provides a separate coordination mechanism, while MIN-Trust restricts
information disclosure across collaborating principals
~\cite{def_agentchain_blockchain_empowered_multi_agent_coor,
def_min_trust_a_minimum_necessary_information_trust}.
Other Set~1 work studies faulty or Byzantine members and changes how collective
evidence is aggregated
~\cite{zheng2026byzantinereliability,
def_blockagents_towards_byzantine_robust_llm_based_m,
def_robust_llm_based_multi_agent_system_with_action}.

D2 also includes dynamic-trust mechanisms such as DynaTrust, A-Trust, and
Ev-Trust, which update influence from interaction history or message signals
~\cite{supp_dynatrust_2026,he2025atrust,
def_ev_trust_an_evolutionarily_stable_trust_mechanis}.
A separate line changes the collective-decision rule itself through local
coordination, Byzantine protocols, conformal social choice, majority-resistant
aggregation, or institutional controls
~\cite{mao2025ibgp,
def_byzantine_fault_tolerant_multi_agent_system_for,
jo2025byzantinerobust,
def_from_debate_to_decision_conformal_social_choice,
consensustrap2026,
def_consensus_driven_metacognition_in_multi_agent_sy,
lee2026robustbyzantine,
def_institutional_ai_governing_llm_collusion_in_mult}.
D2 therefore intervenes by changing communication, contributor influence, or
aggregation. Its guarantees depend on assumptions about identity, membership,
fault bounds, observability, and sometimes synchrony. Reconfiguring trust or
topology can limit future influence, but does not by itself remove manipulated
evidence already committed to persistent state or collective decisions.

\noindent\textbf{D3. Detection, attribution, and containment.}
D3 observes harmful influence and intervenes once suspicious behavior or
propagation is identified. It applies across several attack paths, with much of
the current evidence centered on propagation, routing, and collective
manipulation (P2--P4). G-Safeguard, BlindGuard, XG-Guard, SentinelAgent,
PeerGuard, GUARDIAN, and SentinelNet use communication structure, message or
response signals, peer history, or anomaly scores to identify harmful
participants
~\cite{wang2025gsafeguard,miao2026blindguard,pan2026xgguard,
def_sentinelagent_graph_based_anomaly_detection_in_m,
supp_peerguard_defending_multi_agent_systems_against_backdoor_attacks_through_mutual_,
zhou2025guardian,feng2026sentinelnet}.

Several recent works consider propagation rather than only suspicious endpoints.
Cross-Layer Semantic Flow Reconstruction, CASPIAN, From Spark to Fire, INFA-Guard, PropGuard,
contribution-based methods, and related cascade defenses study how harmful
influence moves across messages, principals, or communication paths
~\cite{def_beyond_input_guardrails_reconstructing_cross_age,
def_caspian_online_detection_and_attribution_of_casc,
def_from_spark_to_fire_modeling_and_mitigating_error,
def_infa_guard_mitigating_malicious_propagation_via,
def_propguard_safeguarding_llm_mas_via_propagation_a,
li2026bpd,
def_memetic_cascade_detection_and_symbolic_immunity}.
Other mechanisms target narrower intervention points, including misinformation
rectification, sentence-level repair, linguistic routing controls,
stubborn-neighbor detection, and game-theoretic or machine-checkable controls
~\cite{li2025argus,e_sentenceRectification2026,
def_linguistic_firewall_geometry_as_defense_in_multi,
abedini2026stubbornneighbors,
def_game_theoretic_multi_agent_control_for_robust_co,
def_collective_hallucination_in_multi_agent_llms_mod,
def_quadsentinel_sequent_safety_for_machine_checkabl,
def_sgto_mas_secure_gorilla_troops_optimization_for,
def_pratyahara_a_neural_tissue_defense_model_for_det}.
SAIGuard moves intervention earlier by predicting the state change caused by a
candidate message and intercepting it before delivery
~\cite{shi2026saiguard}. More broadly, D3 ranges from local message or peer
history to graph- and trace-level observation, with guarantees bounded by the
coverage and integrity of that telemetry
~\cite{pan2026xgguard,feng2026sentinelnet}. Detection and containment do not
imply recovery: once P2 has changed downstream state, stopping further spread
does not undo earlier effects.

\noindent\textbf{D4. State, provenance, and information-flow governance.}
D4 controls what information and derived state may cross principal boundaries,
where they may persist, and what security context remains attached to them. It
therefore most directly addresses propagation, leakage, and delegation
(P2/P6/P7). Set~1 evidence includes differential-privacy and privacy-preserving
systems that limit disclosure across collaborating principals, together with
provenance-aware work on cross-agent information flows
~\cite{def_capri_dp_a_differentially_private_extension_to_c,
goel2026securityprivacy,
def_privacy_preserving_llm_infrastructure_with_multi,
def_cross_agent_multimodal_provenance_aware_framewor}.
Set~1 also includes memory-oriented defenses such as Self-Healing Memory
Architectures
~\cite{def_self_healing_memory_architectures_for_large_lang}.

Set~2 work targets additional forms of shared and latent state, including
Governed Shared Memory and EquiMem
~\cite{def_governed_shared_memory_for_multi_agent_llm_syste,
def_equimem_calibrating_shared_memory_in_multi_agent}.
PRISM models leakage that accumulates across multi-stage generation and
intervenes before a complete secret is emitted, while LCGuard targets
information carried through shared latent or KV state
~\cite{tapwal2026prism,
def_lcguard_latent_communication_guard_for_safe_kv_s}.
Maris applies reference monitors and checked policies to declared cross-principal
message and tool flows, while semantic-taint mechanisms attach tracking
information to derived data
~\cite{cui2025maris,
def_semantic_taint_propagation_embedding_based_seman}.
Other work combines information governance with prompt-injection controls,
verifiable coordination, or contextual privacy mechanisms
~\cite{def_prompt_injection_mitigation_with_agentic_ai_nest,
e_trustworthyAgenticPI2025,
def_dao_agent_zero_knowledge_verified_incentives_for,
arxiv_2508_07667}.
D4's guarantees are strongest when the relevant flow is mediated and its
security context remains trustworthy. Maris, for example, assumes relevant
message and tool flows pass through its reference monitors
~\cite{cui2025maris}, while PRISM and latent-state defenses cover different
leakage channels
~\cite{tapwal2026prism,
def_lcguard_latent_communication_guard_for_safe_kv_s}.
Semantic rewriting, covert channels, and unmediated side effects remain harder
to govern, and recovery requires tracing state that has already been copied or
transformed.

\noindent\textbf{Adaptive defense allocation.}
The four primitives above describe what a defense controls. A smaller emerging
literature asks where limited defensive capacity should be concentrated.
SafeSieve progressively prunes communication, while MESA prioritizes
communication edges according to estimated security relevance
~\cite{
supp_safesieve_from_heuristics_to_experience_in_progressive_pruning_for_llm_based_mul,
e_mesa2026}.
Adjacent adaptive-hardening work uses randomized smoothing or adversarial
co-evolution to improve robustness
~\cite{
def_enhancing_robustness_of_llm_driven_multi_agent_s,
e_advEvoMARL2025}.
These approaches can concentrate stronger protection on interactions that
dominate propagation (P2), collective influence (P4), or downstream authority
(P7), rather than applying the same control everywhere.
We treat adaptive allocation as a cross-cutting strategy rather than a fifth
control primitive because it selects where existing controls should be
strengthened rather than defining a distinct object of enforcement. Its smaller
evidence base also leaves open how well allocation signals transfer across
tasks, models, topologies, and attack strategies.

\subsection{Cross-Cutting Findings}
\label{sec:defense-gaps}

Comparing the defense literature through the contract above reveals three
cross-cutting limitations that are not specific to any one control primitive.

First, \emph{local protection does not imply path closure.}
Attack paths can compose: an attack may enter through P1, persist through P2,
and gain authority through P7. A defense that blocks one transition therefore
does not establish end-to-end protection when another route remains reachable.
The mappings in \cref{tab:defense-family-crosswalk} identify where a control may
interrupt a path, not whether it closes every route to the protected effect.

Second, \emph{defense effectiveness depends on where and when control is applied.}
Some defenses constrain the system before execution, while runtime defenses must
observe enough of the relevant execution to support their intervention.
A global observer may cover more of the execution but rely on coarse or trusted
telemetry, while a local observer may see detailed behavior along only a small
part of the path
~\cite{feng2026sentinelnet,pan2026xgguard}.
Timing also determines what can still be prevented or reversed: ResMAS changes
interaction structure before execution, SAIGuard can block candidate messages
before delivery, while other mechanisms intervene after suspicious influence
has been observed
~\cite{zhou2026resmas,shi2026saiguard,pan2026xgguard}.
Defense guarantees therefore depend on the intervention point and, for runtime
mechanisms, on whether their observation and control are sufficient for the
claimed protection, together with the trust assumptions on which those
mechanisms rely.

Finally, \emph{containment is not recovery.}
Removing the source or stopping further propagation does not restore
the system once its effects have been copied into memory, artifacts,
credentials, delegated authority, or spawned principals
~\cite{liu2026topologymemory,agentworm2026,childinherits2026}.
Recovery requires identifying these descendants, revoking inherited authority,
repairing state, and verifying that the protected property has been
restored. In our map, recovery is evaluated less directly than prevention and
containment.

Together, these findings suggest that end-to-end MAS defense requires more than
a strong local control: defenses must close the relevant attack paths, place
interventions at points where they can affect those paths, provide adequate
observation when runtime decisions depend on it, and account for residual
effects that survive the original compromise.
     
\section{Benchmarks and Evaluation}
\label{sec:benchmarks}







Beyond mapping the attack and defense landscape, we examine how the security of multi-agent systems is evaluated. We first systematize existing benchmarks and their accompanying artifacts. We then identify four gaps in current evaluation practice and discuss the corresponding open challenges.

\subsection{Existing Benchmarks and Artifacts}
\label{sec:existing-benchmarks}

Within the 197-work corpus, we identify 44 works whose main contribution is the security evaluation of multi-agent systems.
These works test attacks, defenses, and security properties under different system designs and threat models. 
We reviewed the paper PDFs for 43 works and the software archive for one Zenodo record without an accompanying paper.
For each work, we record its adversary position A1–A4, configuration dimensions C1–C8, interaction interfaces I1–I6, attack paths P1–P8, and system-level risks R1–R7. We also record their baselines, metrics, observation depth, and artifact availability. All counts below cover the full set of 44 works; the detailed
per-work mapping is provided in \cref{tab:benchmark-audit}.

The works provide broad coverage of both attacks and defenses. Forty-two evaluate an attack or failure, including 16 that also evaluate a defense. CalBench and MAGPIE evaluate privacy without an explicit attacker~\cite{zou2026calbench,juneja2025magpie}. The benchmarks also cover many MAS designs. All 44 report or vary principal composition (C3), 39 cover coordination (C4), and 38 cover communication topology (C1). Membership is fixed in 42 works, while two allow limited runtime spawning. Member adversaries (A3), message transfer (I2), detection (I6), peer or state steering (P5), and collective-decision failure (R3) are the most common settings~\cite{yu2025netsafe,liu2026topologymemory,an2026aciarena,kavathekar2026tamas}.

The evaluations provide different depths of execution evidence. Classified by their deepest observation, 20 use full traces, 20 use principal-level messages or actions, two use episode outcomes, and two use final outputs. Thirty-one works identify code, data, logs, a test tool, or a project page, including 26 with an exact URL. The works also report communication topology for 38 works, model information for 32, and an exact principal count for 18.

\begin{benchmarksummary}
\textbf{Summary.} The 44 works evaluate diverse MAS designs and security questions, with varied observation depth and artifact support.
\end{benchmarksummary}

\subsection{Gaps and Open Challenges}
\label{sec:evaluation-gaps}

Our analysis identifies four gaps in current evaluation practice. First, evaluations often show that a failure occurs without isolating how multi-agent interaction affects it. Second, metric names may hide what is measured, while execution traces rarely support causal diagnosis. Third, comparisons within one benchmark do not by themselves provide configurable or extensible components for other MAS designs. Finally, current evaluations largely focus on fixed, closed systems rather than practical open settings.

\subsubsection{Isolating Multi-Agent Effects}
\label{sec:gap-counterfactual}

Determining how interaction changes a security outcome remains an open evaluation challenge.
 As defined in Section~\ref{sec:scope}, a failure may remain unchanged without interaction, become worse because of interaction, appear only because of interaction, or involve a security property that exists only among multiple principals.
 Distinguishing these four roles requires comparisons designed for the specific role being tested.

When an attack has a single-principal counterpart, a matched single-principal comparison can distinguish an inherited effect from an interaction-amplified effect. Our audit identifies 21 works that evaluate at least one of these two effects. Six compare an MAS with a single-principal execution~\cite{arora2026safeagents,eval_harnessaudit2026,e_openclawSecurity2026,zhao2026macbench,elyagoubi2026agentleak,s1collabjailbreak2026}. The remaining 15 evaluate the attack in an MAS or compare different MAS configurations. These evaluations show that the attack succeeds under the tested conditions, but they do not determine whether the same failure would occur with one principal or become worse through interaction.

When no meaningful single-principal counterpart exists, the evaluation should preserve the MAS and vary the relation under study. Among 36 such works, 12 include a relation-focused comparison. The remaining 24 use another control or provide no clear relation-focused comparison. For example, MESA removes or masks communication edges, while CalBench compares models and coordination protocols under the same private-information constraints~\cite{e_mesa2026,zou2026calbench}.

\begin{benchmarksummary}
\textbf{Summary.} Interaction-specific comparison remains an open evaluation challenge. Future benchmarks should match their comparison design to the interaction effect being tested.
\end{benchmarksummary}

\subsubsection{Comparable and Diagnostic Metrics}
\label{sec:gap-metrics}

Current metrics are inadequate in two respects: their outcome-level measures are not comparable across benchmarks, and their trace-level measures do not explain how a compromise propagates across agents.

\paragraphtitle{Underspecified ASR.} Attack Success Rate (ASR) is the most common security metric in our corpus. Thirteen of the 44 works compute ASR in their own evaluation, while ATAG additionally records ASR values reported by other sources~\cite{supp_atag_ai_agent_application_threat_assessment_with_attack_graphs}. Across these studies, ASR refers to different evaluation units and success conditions because the benchmarks study different attacks. The name “ASR” alone therefore does not identify what event is counted or over which eligible population. An ASR of 60\% in one benchmark may therefore measure a different event from an ASR of 60\% in another. The values cannot be directly compared, ranked, or aggregated unless these definitions match. When propagation is part of the claim, an end-to-end ASR also cannot distinguish an attack stopped at entry from one that spreads across several agents before being blocked.

Future benchmarks should qualify ASR by its evaluation unit, using names such as per-principal ASR, per-message ASR, or per-run ASR, and define the counted success event. Each result should provide its taxonomy mapping, evaluation unit, execution protocol, and verification method.
When propagation is part of the claim, stage-level measures should follow the relevant I1–I6 path to the verified R1–R7 consequence.
These stages should be selected based on the attack rather than imposed on every risk. Existing work already measures individual dimensions such as harm amplification, topology-dependent leakage, and internal-channel exposure~\cite{rahman2026harp,liu2026topologymemory,elyagoubi2026agentleak}. Benign task utility should be reported separately from security outcomes.

\paragraphtitle{Missing Causal Provenance.} Full execution traces provide more information than final-output evaluation, but they do not necessarily explain why a failure occurred. Twenty records use the full execution trace as their deepest observation. These traces may show that one principal received malicious information, another updated memory, and a third invoked a tool, without showing how these events depend on one another. Across the corpus, ATAG models attack dependencies and Agent-BOM reconstructs dependencies from executions~\cite{supp_atag_ai_agent_application_threat_assessment_with_attack_graphs,eval_securityauditable2026}. Agent-BOM is the only one of these 20 full-trace records coded as reporting an explicit dependency or provenance graph. This evidence is unnecessary for benchmarks that report only final outcomes, but it is needed for claims about propagation, responsibility, or defense placement.

Benchmarks that make such claims should represent the relevant execution path as a provenance graph. Its nodes should identify the principals and events at the I1--I6 interfaces, including attacker-controlled inputs, messages, memory operations, tool results, delegation decisions, and security-sensitive actions. Its edges should record data, control, and delegation dependencies. The path should begin with the first adversarially influenced event or violated trust boundary and end with a verified R1--R7 consequence. Because a dependency graph does not prove causation, the claimed path should also be tested through controlled replay or ablation.

\begin{benchmarksummary}
\textbf{Summary.} Thirteen works compute ASR in their own evaluation, while ATAG additionally records ASR values reported by other sources. Twenty records use the full execution trace as their deepest observation, and Agent-BOM is the only one coded as reporting an explicit dependency or provenance graph.
\end{benchmarksummary}

\subsubsection{Configurable and Extensible Benchmarks}
\label{sec}

Thirty-nine of the 44 works compare two or more evaluation conditions. These comparisons include component or attack controls, defense controls, single-principal baselines, and relation-focused MAS comparisons. They support controlled variation within each study, but do not show whether benchmark factors can be changed independently or reused elsewhere. We call a benchmark \emph{configurable} when its supplied factors can be changed independently and \emph{extensible} when new components can be added through a documented interface.

ACIARena defines interfaces for MAS implementations, attacks, defenses, and evaluation modules~\cite{an2026aciarena}. RiskLab provides configurable experiment components and a registry for adding risk detectors~\cite{jiang2026risklab}. A2ASecBench uses an LLM-based adapter to apply attack descriptions to different A2A scenarios~\cite{li2026a2asecbench}. These benchmarks expose useful extension points, but their task, trace, attack, and verifier formats differ. Transferring a component between them would therefore require an additional adapter. Similarly, the 31 works with identified artifacts may provide access to code or data without making their components reusable. Because our audit does not systematically record interface reuse, we do not quantify how often such reuse occurs.

Benchmarks intended for cross-system comparison should expose the factors named in their claims through documented interfaces. A paired test should change one factor while preserving the protected property and outcome definition. Differences between system interfaces should be handled through validated adapters. Each run should record four groups of information: the model specification, execution settings, environment state, and verifier version.

\begin{benchmarksummary}
\textbf{Summary.} Thirty-nine works compare conditions within their own benchmark, but these comparisons do not by themselves establish configurability or extensibility. Future cross-system benchmarks should use documented interfaces while preserving the same security goal and success condition.
\end{benchmarksummary}

\subsubsection{Open-System Evaluation}
\label{sec:gap-realism}
Current benchmarks primarily evaluate closed systems whose members and trust relationships are fixed before execution. Forty-two of the 44 works use fixed membership, while the other two allow limited runtime spawning under one administrator. Ten works use deception games or negotiation environments, and many others rely on synthetic tasks with predefined roles. These settings support controlled experiments but do not capture how practical MAS operate across changing trust boundaries.

Identity, trust, authority, and policy also matter in closed systems. However, they become more dynamic and consequential when previously unknown principals interact across administrative boundaries~\cite{ko2026sevenchallenges,schroederdewitt2025openchallenges}. Current benchmarks largely begin after these relationships have been established. They therefore provide little evidence about how principals enter the system, how trust changes during execution, or how access is removed.

Future benchmarks should evaluate deployed coordination protocols and frameworks with independently managed principals. Testbeds should include heterogeneous implementations, persistent state, external services, and changing membership. They should cover the lifecycle from discovery and admission through operation and removal. Security outcomes should reflect consequences across real trust boundaries rather than only success or failure in a synthetic task.

\begin{benchmarksummary}
\textbf{Summary.} Forty-two works use fixed membership, while two allow limited spawning under one administrator. Evaluating changing membership and cross-domain trust remains an open benchmark problem.
\end{benchmarksummary}

\section{Conclusion}\label{sec:conclusion}

This SoK frames multi-agent LLM security around end-to-end execution. Across
197 works, we organize MAS security through system configuration, interaction
interfaces, attack paths, defenses, and evaluation. Our analysis shows that
security depends on how adversarial influence crosses principal boundaries,
what state and authority remain reachable, whether defenses can close the path,
and whether evaluations verify the resulting system-level effect. This
execution-centered view provides a common basis for comparing security claims
across MAS designs.

\clearpage
\section*{Ethical Considerations}

This SoK synthesizes publicly available research on the security of multi-agent
systems. It does not involve human-subject studies, private user data, or
experiments against live or third-party systems. Our analysis is based on
published papers, public artifacts, advisories, and bibliographic metadata.
The topic is inherently dual use: organizing attack mechanisms and system
weaknesses can help both defenders and attackers understand where multi-agent
systems fail. We therefore emphasize security properties, interaction paths,
threat assumptions, defenses, and evaluation evidence rather than operational
payloads or step-by-step exploitation procedures. We do not disclose
non-public vulnerabilities or unnecessary identifying details.
Finally, we distinguish experimental evidence from deployment prevalence.
A vulnerability demonstrated in a benchmark or controlled environment does not
by itself imply that the same failure is common in deployed systems. We report
claims within the scope and assumptions supported by the underlying studies.

\section*{Open Science}

The literature corpus and associated artifacts will be released publicly with the final version of the paper.


\section*{Acknowledgement}
Research reported in this publication was supported by an Amazon Research Award, Fall 2025, and by the 2026 Amazon Nova AI Challenge: Trusted Software Agents. The views and conclusions contained herein are those of the authors and should not be interpreted as necessarily representing the official policies of the supporting sponsors.

{\footnotesize
\bibliographystyle{plain}
\bibliography{refs}
}


\appendix
\section{Evaluation Audit Details}
\label{app:evaluation-audit}

Table~\ref{tab:benchmark-audit} provides the detailed mapping for the 44 works in our
evaluation audit. It records their A, C, I, P, and R assignments together with
observation depth, comparison design, membership setting, artifact availability,
and reported system information. The taxonomy codes follow Sections~2--4.

\begin{sidewaystable*}[p]
\centering
\fontsize{6.5}{7.4}\selectfont
\setlength{\tabcolsep}{1.5pt}
\renewcommand{\arraystretch}{1.02}
\begin{tabular*}{\textheight}{@{\extracolsep{\fill}}p{3.25cm}cp{1.15cm}p{2.15cm}p{1.75cm}p{1.95cm}p{1.75cm}llclc@{}}
\toprule
\textbf{Work} & \textbf{Role} & \textbf{Adv.} & \textbf{Config.} & \textbf{Interface} & \textbf{Path} & \textbf{Risk} & \textbf{Obs.} & \textbf{Comp.} & \textbf{Member} & \textbf{Artifact} & \textbf{Report} \\
\midrule
A2ASecBench~\cite{li2026a2asecbench} & A+D & A1--A4 & C1--C8 & I1--I6 & P1--P8 & R1,R3--R7 & Trace & AC & Fixed & URL & MT \\
Achilles Heel~\cite{eval_achilles2025} & A & A3 & C1,C3--C7 & I2,I4,I5 & P2,P4,P5,P7,P8 & R1--R7 & Episode & Rel & Fixed & -- & MT \\
ACIARena~\cite{an2026aciarena} & A+D & A2--A4 & C1--C3,C5,C6,C8 & I1--I3,I5,I6 & P1--P3,P5,P6,P8 & R1,R3--R5,R7 & Episode & Rel & Fixed & URL & MT \\
Agent-BOM~\cite{eval_securityauditable2026} & A & A1,A3 & C2,C3,C5--C8 & I1--I4,I6 & P1--P3,P5--P8 & R1,R4--R7 & Trace & None & Spawn & -- & M \\
AgentLeak~\cite{elyagoubi2026agentleak} & A & A1--A4 & C1--C8 & I1--I4,I6 & P1--P3,P5--P8 & R1,R5 & Trace & SP & Fixed & URL & NMT \\
AgentXposed~\cite{xie2025whosthemole} & A+D & A3 & C1,C3,C4,C6,C8 & I2,I5,I6 & P2,P4,P5,P8 & R1--R4,R7 & Events & AC & Fixed & URL & NMT \\
Algorithmic Cowardice~\cite{yankeloviz2026cowardice} & A & A3 & C1--C4,C6--C8 & I2,I6 & P4,P5 & R1,R4 & Events & AC & Fixed & URL & NMT \\
Among Us~\cite{milkowski2026amongus} & A+D & A3 & C1,C3,C4,C6,C8 & I2,I5,I6 & P4,P5 & R2,R3 & Events & AC & Fixed & URL & MT \\
ATAG~\cite{supp_atag_ai_agent_application_threat_assessment_with_attack_graphs} & A & A1,A2 & C1--C3,C5,C7,C8 & I1--I5 & P1--P3,P5,P7,P8 & R1,R4--R6 & Events & None & Fixed & URL & MT \\
BAD-ACTS~\cite{nother2026badacts} & A+D & A1--A4 & C1--C8 & I1--I6 & P1--P8 & R1--R7 & Trace & AC & Fixed & URL & MT \\
CalBench~\cite{zou2026calbench} & S & -- & C1--C8 & I2--I6 & P3--P8 & R2,R3,R5,R7 & Trace & Rel & Fixed & Claim & NMT \\
Capability Paradox~\cite{eval_capabilityparadox2026} & A & A2 & C1--C4,C7,C8 & I1,I2,I4,I6 & P1,P2,P5,P7 & R1,R4,R6 & Trace & Rel & Fixed & -- & MT \\
Child Inherits~\cite{childinherits2026} & A & A2,A3 & C1,C3,C5--C8 & I1,I3,I4,I6 & P1--P3,P5--P8 & R1,R4--R7 & Final & Rel & Spawn & -- & T \\
Collab. Jailbreaking~\cite{s1collabjailbreak2026} & A & A2 & C1--C8 & I1--I6 & P1--P5,P7 & R1,R3,R4,R6 & Events & SP & Fixed & -- & MT \\
Collusion Interpretability~\cite{e_collusionInterpretability2026} & A & A3 & C1--C5,C8 & I2--I6 & P2,P4,P5,P7 & R2--R4,R6 & Events & AC & Fixed & URL & NT \\
Colosseum~\cite{nakamura2026colosseum} & A+D & A3 & C1--C5,C7,C8 & I2,I4--I6 & P2--P5 & R1--R3 & Trace & Rel & Fixed & URL & MT \\
CRDA~\cite{eval_crda2024} & A & A3 & C2--C6 & I2,I3,I5,I6 & P2--P5 & R1,R3,R4 & Events & Rel & Fixed & -- & N \\
Defection~\cite{eval_defection2026} & A & A3 & C2--C8 & I2--I6 & P3--P5,P7,P8 & R2--R4,R7 & Events & AC & Fixed & -- & N \\
GAMBIT~\cite{lemercier2026gambit} & A+D & A3 & C1--C4,C6,C8 & I2,I5,I6 & P4,P5 & R2,R3 & Events & SP & Fixed & URL & MT \\
GAMMAF~\cite{mateotorrejon2026gammaf} & A+D & A3 & C1--C4,C6--C8 & I2,I5,I6 & P2--P5,P8 & R1--R3,R7 & Events & AC & Fixed & URL & NMT \\
Harness Audit~\cite{eval_harnessaudit2026} & A & A1 & C1--C5,C7,C8 & I1--I6 & P1,P2,P6--P8 & R1,R4--R6 & Trace & SP & Fixed & URL & MT \\
HARP~\cite{rahman2026harp} & A+D & A3,A4 & C1--C5,C7,C8 & I2--I6 & P2,P4,P5,P7 & R1,R3,R4,R6 & Trace & AC & Fixed & -- & NMT \\
Hidden in Plain Text~\cite{mathew2025hidden} & A+D & A3 & C2--C4,C6,C8 & I2,I5,I6 & P4,P6 & R2,R3,R5 & Events & DC & Fixed & -- & -- \\
LieCraft~\cite{olson2026liecraft} & A+D & A3 & C1,C3,C4,C6,C8 & I2,I5,I6 & P4,P5 & R2,R3 & Events & AC & Fixed & URL & NMT \\
LLM Drift~\cite{e_llmDrift2026} & A & A3 & C1--C8 & I2,I3,I5,I6 & P2--P5 & R2--R4 & Trace & None & Fixed & URL & MT \\
LLM-Deliberation~\cite{eval_coopmalicious2024} & A & A3 & C1--C7 & I2,I3,I5,I6 & P3--P6 & R2--R5 & Events & Rel & Fixed & Claim & MT \\
MAC-Bench~\cite{zhao2026macbench} & A+D & A2,A3 & C1--C5,C7,C8 & I1--I6 & P1,P3--P8 & R3--R7 & Trace & SP & Fixed & Claim & MT \\
MAGPIE~\cite{juneja2025magpie} & S & -- & C1--C8 & I2,I3,I5,I6 & P3--P6 & R2--R5 & Trace & DC & Fixed & URL & NMT \\
MAMA~\cite{liu2026topologymemory} & A & A3 & C1--C3,C5,C6,C8 & I2,I3,I6 & P2,P3,P5,P6 & R1,R5 & Events & Rel & Fixed & URL & NMT \\
MedSentry~\cite{chen2025medsentry} & A+D & A2,A3 & C1--C6,C8 & I1,I2,I5,I6 & P1,P2,P4--P6 & R1,R3,R4 & Events & Rel & Fixed & Claim & NMT \\
MESA~\cite{e_mesa2026} & A & A4 & C1--C4,C7,C8 & I2,I5,I6 & P2,P3,P5 & R1,R3 & Events & Rel & Fixed & URL & T \\
NetSafe~\cite{yu2025netsafe} & A & A3 & C1--C4,C6,C8 & I2,I5,I6 & P2--P5 & R1,R3 & Events & Rel & Fixed & URL & NT \\
OpenClaw Security~\cite{e_openclawSecurity2026} & A & A1,A4 & C1--C5,C7,C8 & I1,I2,I4--I6 & P1,P2,P7,P8 & R1,R4,R6 & Trace & SP & Fixed & -- & NMT \\
PEAR~\cite{dong2026pear} & A & A2,A4 & C2--C5,C7,C8 & I1--I4,I6 & P1--P3,P5--P8 & R4--R7 & Trace & Rel & Fixed & Claim & NM \\
Privacy Leakage~\cite{eval_privacyleakage2025} & A & A3 & C1--C6,C8 & I1--I4,I6 & P1,P2,P4--P7 & R1,R2,R5,R6 & Events & SP & Fixed & -- & NMT \\
RiskLab~\cite{jiang2026risklab} & A+D & A3 & C1--C5,C8 & I2--I6 & P4,P5 & R2--R4 & Trace & None & Fixed & URL & MT \\
SafeAgents~\cite{arora2026safeagents} & A & A1,A2 & C1--C4,C7,C8 & I1,I2,I4--I6 & P1,P2,P5,P7 & R4,R6 & Trace & SP & Fixed & URL & MT \\
Security Case Study~\cite{fan2025securitycase} & A & A2--A4 & C1--C8 & I1--I6 & P1--P5,P7,P8 & R1,R3,R4,R6,R7 & Events & DC & Fixed & Up & T \\
SNEAK~\cite{eval_sneak2026} & A & A3 & C2--C4,C8 & I2,I5,I6 & P6 & R5 & Events & AC & Fixed & -- & -- \\
TAMAS~\cite{kavathekar2026tamas} & A+D & A1--A3 & C1--C4,C6--C8 & I1,I2,I4--I6 & P1,P2,P4,P5,P7,P8 & R1--R4,R6,R7 & Trace & Rel & Fixed & URL & NMT \\
Terrarium~\cite{eval_terrarium2025} & A & A3,A4 & C1--C8 & I1--I6 & P2--P6,P8 & R1--R5,R7 & Trace & AC & Fixed & URL & T \\
TrinityGuard~\cite{wang2026trinityguard} & A & A1--A3 & C1--C8 & I1--I6 & P1--P8 & R1--R7 & Trace & None & Fixed & URL & T \\
Trust Paradox~\cite{xu2026trustparadox} & A+D & A3 & C1--C4,C6--C8 & I2,I4--I6 & P3--P7 & R5,R6 & Final & Rel & Fixed & Claim & MT \\
WOLF~\cite{eval_wolf2025} & A & A3 & C1,C3,C4,C6,C8 & I2,I3,I5,I6 & P3--P5 & R2,R3 & Trace & AC & Fixed & URL & T \\
\bottomrule
\end{tabular*}
\caption{Properties of the 44 security-evaluation works. Role gives attack or failure evaluation (A), defense evaluation (D), and security-property evaluation without an explicit attacker (S). Obs. is the deepest data consumed by the evaluation: final output, episode outcome, principal or message events, or a full trace. Comp. gives the strongest coded comparison: attack or component control (AC), defense or component control (DC), matched single-principal comparison (SP), relation-focused MAS comparison (Rel), or none. Member reports fixed membership or limited runtime spawning. Artifact distinguishes an exact evaluated-artifact URL (URL), a claimed artifact without a recoverable URL (Claim), an upstream-only URL (Up), and no identified artifact (--). Report marks an exact principal count (N), model information (M), and communication topology (T). Taxonomy codes follow Sections~2--4.}
\label{tab:benchmark-audit}
\end{sidewaystable*}

\end{document}